\documentclass[twocolumn]{aastex631}

\shorttitle{A Stable Coronal Spectrum on XZ Tau}
\shortauthors{Silverberg et al.}
\graphicspath{{./}{figures/}}
\begin{document}

\title{Stable Coronal X-Ray Emission Over Twenty Years of XZ Tau}

\author[0000-0002-0786-7307]{Steven M. Silverberg}
\affiliation{Smithsonian Astrophysical Observatory, MS 70, 60 Garden St., Cambridge, MA 02138}

\author[0000-0003-4243-2840]{Hans Moritz G\"unther}
\affiliation{MIT Kavli Institute for Astrophysics and Space Research, 77 Massachusetts Avenue, Cambridge, MA 02139, USA}

\author[0000-0002-1131-3059]{Pragati Pradhan}
\affiliation{Embry-Riddle Aeronautical University, Prescott, AZ, USA}

\author[0000-0002-7939-377X]{David A. Principe}
\affiliation{MIT Kavli Institute for Astrophysics and Space Research, 77 Massachusetts Avenue, Cambridge, MA 02139, USA}

\author[0000-0002-5094-2245]{P. C. Schneider}
\affiliation{Hamburger Sternwarte, Gojenbergsweg 112, D-21029, Hamburg, Germany}

\author[0000-0002-0826-9261]{Scott J. Wolk}
\affiliation{Smithsonian Astrophysical Observatory, MS 70, 60 Garden St., Cambridge, MA 02138}



\begin{abstract}
XZ Tau AB is a frequently observed binary YSO in the Taurus Molecular Cloud; XZ Tau B has been classified as an EXOr object. We present new \textit{Chandra}/HETG-ACIS-S observations of XZ Tau AB, complemented with variability monitoring of the system with \textit{XMM-Newton}, to constrain the variability of this system and identify high-resolution line diagnostics to better understand the underlying mechanisms that produce the X-rays. We observe two flares with \textit{XMM-Newton}, but find that outside of these flares the coronal X-ray spectrum of XZ Tau AB is consistent over twenty years of observations. We compare the ensemble of XZ Tau X-ray observations over time 
with the scatter across stars observed in point-in-time observations of the Orion Nebula Cluster and find that both overlap in terms of plasma properties, i.e., some of the scatter observed in the X-ray properties of stellar ensembles stems from intrinsic source variability.
\end{abstract}

\keywords{Young star, X-ray, flare star Facilities: Chandra, XMM-Newton
}


\section{Introduction} \label{sec:intro}

Stars and their circumstellar disks form as a result of gravitational contraction of molecular clouds. The early stages of low-mass ($<2 M_{\odot}$) stellar evolution are characterized by violent accretion events, large molecular outflows and jets as star-disk systems clear their surrounding environment. A subset of pre-MS stars undergoing this violent accretion are part of an observationally rare but important class called FUor and EXor (after their namesakes FU Ori and EX Lup). These objects undergo extreme mass accretion events ($\Delta \mathbf{\dot{M}} \sim 10^{2-4}$) where their optical brightness can increase by $\Delta V \sim 3-5$ \citep{2014prpl.conf..387A}. FUor and EXor objects are distinguished by their outburst intensity and frequency: FUors exhibit brighter $\Delta V \sim 3-5$~mag events which last years-centuries, while EXors undergo less intense but more frequent outbursts on timescales of months-years. While it is clear FUor/EXors drastically increase in magnitude as a result of sudden accretion, it is debated as to why these objects undergo these intense mass accretion events. Several theories have been proposed including: disk fragmentation \citep{2013A&A...552A.129V} and/or perturbations from a binary component or massive planet causing disk instabilities \citep{1990MNRAS.242..439C} or even the ``consumption'' of tidally disrupted protoplanets.

X-ray emission is ubiquitous among low-mass pre-MS stars due to their convective zones and fast rotational periods which generate strong magnetic dynamos. In particular, several distinct physical mechanisms are capable of producing X-rays in pre-MS stars: magnetically heated coronae with characteristic temperatures of $\sim0.1-10.0$ keV that result in continuum and line emission dependent on coronal abundances \citep[e.g.][]{2005ApJS..160..401P,2007A&A...473..589S}, shock-heated plasma (at characteristic temperatures of $\sim0.3$ keV) as a result of mass funneled from the circumstellar disk along magnetic field lines and onto the star \citep[see review by][]{2022arXiv220706886S}, and star-disk magnetic reconnection events which can magnetically heat material to temperatures in excess of $\sim8.6$ keV \citep{2005ApJS..160..469F}. A multiwavelength campaign to study the outburst of FUor/EXor type object V1647 Ori detected a $\sim2$ order of magnitude increase in X-ray flux associated with star-disk magnetospheric interactions \citep{2004Natur.430..429K}. A handful of cases have also been identified where angular momentum is lost as jets launched from the system during the Class II stage, which are also seen in X-rays \citep{2001Natur.413..708P,Favata_2002,2011A&A...530A.123S}, mostly from plasma with $kT\approx0.3$\,keV \citep[e.g.][]{2006A&A...448L..29G,2007A&A...462..645B}. This X-ray emission likely comes from shock-heated material traveling away from the source with velocities in excess of 300\,km\,s$^{-1}$  \citep[see, e.g., DG Tau, ][]{2008A&A...478..797G, 2008A&A...488L..13S}. X-ray observations of FUor/EXor type objects have revealed indications of X-ray emitting jets \citep[Z Cma]{2009A&A...499..529S}, X-ray bright accretion hot spots \citep{2012ApJ...754...32H}, and magnetic reconnection events \citep{2004Natur.430..429K}.

Eventually, pre-MS stars clear out their surrounding molecular envelopes and disks and thus lose many of the above physical mechanisms capable of producing X-rays even while they remain coronally active. Therefore, it is imperative to investigate features during this young, embedded stage of stellar evolution to understand how stars evolve and, in particular, investigate what impact these sources of X-rays have on circumstellar disks and the eventual formation of planets \citep{2011MNRAS.412...13O,2013ApJ...765....3S,2013ApJ...772....5C}. While numerous studies have been published revealing a wealth of data regarding star formation \citep{2005ApJS..160..319G,2007AandA...468..353G}, the vast majority of the observational work has been limited to single snapshots in time for a variety of objects, rather than tracking particular sources over longer time periods. One potential source to analyze over this longer time frame is the YSO binary system XZ Tau.

% The Chandra High Energy Transmission Gratings (HETG) combined with ACIS-S provides spectral resolution capable of resolving ratios of temperature sensitive lines of H-like and He-like ions, as well as the density sensitive forbidden to inter combination line intensities within the triplet line complexes of He-like ions \citep[e.g. Mg XI, Ne IX, OVII][]{2007ApJ...671..592H}. HETG observations of the star- disk system TW Hya revealed a soft ($\sim0.26$ keV) plasma with high density characteristic of an accretion shock onto the star \citep{2002ApJ...567..434K,2010ApJ...710.1835B}. HETG is the superior instrument for studying the high energy plasma states crucial for understanding accretion and coronae in this short-lived stage. 

\subsection{The XZ Tau AB system\label{sect:xzt}}
XZ Tau AB\footnote{For clarity, we refer to the combined binary system as XZ Tau AB, while each individual component is referred to as XZ Tau A or XZ Tau B, respectively.} is a close separation binary with $a \sim 0.3''$ or 42 au \citep{1990A&A...230L...1H} composed of an M1 and an M2 pre-MS star. Each member hosts its own circumstellar disk \citep{2015ApJ...811L...4Z}, and their small separation likely induces disk disruptions leading to mass accretion events. Optical spectroscopy resolving the binary has demonstrated that each member is a highly accreting source \citep{2001ApJ...556..265W}. Moreover, the spectrum of XZ Tau B is heavily veiled and shows spectral features similar to that of DG Tau, a star-disk system with intense accretion and an observed jet resolved at both optical and X-ray wavelengths \citep{2008A&A...478..797G}. Both XZ Tau A and XZ Tau B exhibit jets at optical wavelengths \citep{2008AJ....136.1980K} while a complex bubble-like system encompasses both stars.

ALMA 1.3mm continuum observations have detected dust emission associated with both circumstellar disks. 12CO emission traces collimated outflows surrounding the XZ Tau A system, typically a signature of youth \citep{2015ApJ...811L...4Z,2006ApJ...646.1070A}. The 1.3mm continuum detection of XZ Tau B indicates an unusually small circumstellar disk with a radius of only 3.6 AU and an inner cavity of 1.3 AU that the authors attribute to ongoing planet formation \citep{2016ApJ...825L..10O}. Such a small disk may be unable to sufficiently shield itself from intense X-ray irradiation and thus the impact on planet formation should be investigated. The classification of XZ Tau B as an EXor object \citep{2004A&A...419..593C,2009ApJ...693.1056L} in combination with its unusually high accretion rate and small disk suggests this pre-MS star may be going through rapid stellar evolution and its disk may not survive for much longer.

\begin{figure*}
    \centering
    \includegraphics[width=0.95\textwidth]{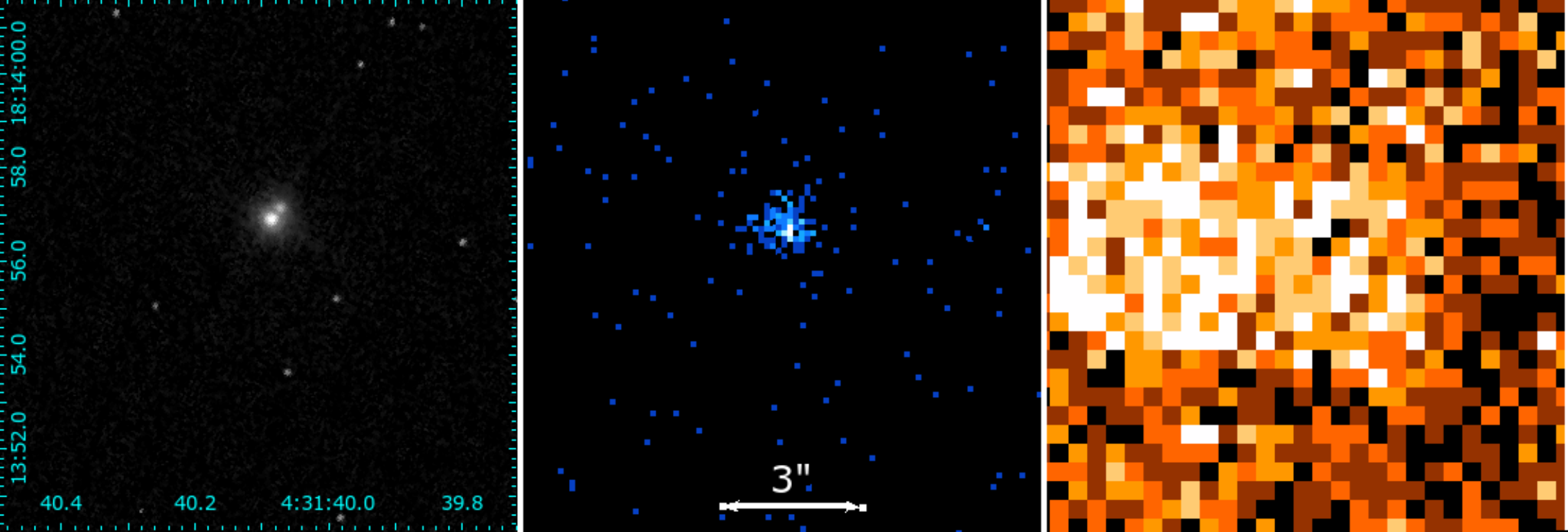}
    \caption{XZ Tau and its environs; all three panels are on the same spatial scale. \textit{Left}: 2004 HST/ACS/F625W image of XZ Tau AB. \textit{Middle}: 2018 \textit{Chandra}/ACIS observations (obsid 21948 shown here) do not resolve XZ Tau AB, but have a tight PSF with low background. \textit{Right}: 2020 \textit{XMM-Newton} observations (obsid 0865040201 shown here) have more counts, but a much wider PSF.}
    \label{fig:HST_comparison}
\end{figure*}

XZ Tau AB has been observed with multiple X-ray instruments between 1990 and the present and displays both long and short term variability. None of these observations are able to resolve the $\sim0.3''$ binary; \textit{Chandra}'s PSF comes close but does not achieve it, while \textit{XMM-Newton} is incapable of it, as seen in Figure \ref{fig:HST_comparison}. \citet{2003A&A...403..187F} reported variability during their $\sim 50$ ks observation of XZ Tau AB. While the count rate increased linearly during the observation, $N_{H}$ decreased from $1.06 \times 10^{22} \mathrm{cm}^{-2}$ to $0.26 \times 10^{22} \mathrm{cm}^{-2}$ while the plasma temperature increased by a similar factor. An emission measure ratio of $\sim{1300}$ between the soft (0.14 keV) and hard (2.29 keV) temperature component at the onset of the count rate increase strongly suggest the presence of either accretion or jet emission. In 2006, a five day monitoring campaign of XZ Tau AB displayed only low levels of $N_{H} \sim 0.1 \times 10^{22} \mathrm{cm}^{-2}$ with $\mathrm{kT_{X1}}, \mathrm{kT_{X2}} = 0.84$ keV and 4.6 keV respectively, although spectral analysis was complicated by extensive background emission during the observation. \citet{2006A&A...453..241G} presented a re-analysis of the \citet{2003A&A...403..187F} data, showing that the variable X-ray spectrum could be fit with either a low or high $N_{H}$ depending on the coronal abundances assumed in the model. However, none of these observations incorporate high-resolution gratings spectra, potentially capable of breaking degeneracies between X-ray emission mechanisms by enabling resolution of temperature- and density-sensitive line ratios \citep[e.g. Mg XI, Ne IX, OVII;][]{2007ApJ...671..592H,2002ApJ...567..434K,2010ApJ...710.1835B}.

In this paper, we present new observations of XZ Tau AB with \textit{Chandra}/HETG, complemented with observations with \textit{XMM-Newton} collected as part of a larger X-ray monitoring campaign on the Taurus star-forming region to constrain variability observed in the \textit{Chandra} observations. Using these data, we analyze the present-day state of XZ Tau AB, in comparison to previous observations. We also present contemporaneous ground-based optical observations to assess whether XZ Tau AB was in outburst during the most recent \textit{Chandra} and \textit{XMM-Newton} observations.

We discuss the observations and data reduction methods in Sect.~\ref{sec:observations}. In Section \ref{sec:analysis} we summarize the characteristics of our observed X-ray data. In Section \ref{sec:discussion} we outline the implications of these observations, and compare our assessments to past work. We summarize our work in Section \ref{sec:conclusion}.

\section{Observations and Data Reduction} \label{sec:observations}

We summarize our new observations in Table \ref{tab:obs_sum}. Below, we briefly describe the observations and data reduction.

\begin{deluxetable*}{lcccc}
\tablecaption{Summary of New Observations of XZ Tau AB \label{tab:obs_sum}}
\tablehead{
\colhead{obsid} & \colhead{Telescope/Instrument} & \colhead{Start (MJD)} & \colhead{Start Date (UTC)} & \colhead{Duration (ks)}}
\startdata
21946      & \textit{Chandra}/ACIS/HETG & 58415.68640 & 2018-10-24 & 12.0 \\
20160      & \textit{Chandra}/ACIS/HETG & 58417.26343 & 2018-10-26 & 41.5 \\
21947      & \textit{Chandra}/ACIS/HETG & 58418.28612 & 2018-10-27 & 12.0 \\
21948      & \textit{Chandra}/ACIS/HETG & 58418.62041 & 2018-10-27 & 56.0 \\
20161      & \textit{Chandra}/ACIS/HETG & 58419.90608 & 2018-10-28 & 48.5 \\
21950      & \textit{Chandra}/ACIS/HETG & 58421.71834 & 2018-10-30 & 14.0 \\
21951      & \textit{Chandra}/ACIS/HETG & 58422.54435 & 2018-10-31 & 36.5 \\
21952      & \textit{Chandra}/ACIS/HETG & 58424.40864 & 2018-11-02 & 12.5 \\
21953      & \textit{Chandra}/ACIS/HETG & 58425.19608 & 2018-11-03 & 36.5 \\
21954      & \textit{Chandra}/ACIS/HETG & 58433.72045 & 2018-11-11 & 25.5 \\
21965      & \textit{Chandra}/ACIS/HETG & 58434.31770 & 2018-11-12 & 25.5 \\
0865040201 & \textit{XMM-Newton}/EPIC   & 59079.67552 & 2020-08-18 & 36.8 \\
0865040301 & \textit{XMM-Newton}/EPIC   & 59083.65674 & 2020-08-22 & 40.0 \\
0865040401 & \textit{XMM-Newton}/EPIC   & 59089.81463 & 2020-08-28 & 33.0 \\
0865040601 & \textit{XMM-Newton}/EPIC   & 59095.94792 & 2020-09-03 & 33.0 \\
0865040701 & \textit{XMM-Newton}/EPIC   & 59104.26799 & 2020-09-12 & 33.0 \\
0865040501 & \textit{XMM-Newton}/EPIC   & 59110.54296 & 2020-09-18 & 47.9 \\
\enddata
\end{deluxetable*}

\subsection{Chandra}

XZ Tau AB was observed by \textit{Chandra} eleven times over the span of three weeks, from 2018 October 24 through 2018 November 12, with the High-Energy Transmission Grating Spectrograph (HETGS)  \citep{2005PASP..117.1144C}. The aimpoint was centered between XZ Tau AB and HL Tau, with the goal of observing both sources in parallel. Data were reduced with the Chandra Interactive Analysis software (CIAO; v.\ 4.14). The observations were energy filtered (0.5-8.0 keV) and time-filtered on good time intervals to reduce flaring particle background. Zeroth-order and gratings spectra were extracted with standard procedures in CIAO.

\subsection{XMM-Newton}

XZ Tau AB was observed by the X-ray Multimirror Mission (\textit{XMM-Newton}) observatory six times over the course of 33 days from 2020 August 18 through 2020 September 19, as part of a larger campaign to monitor variability of young stellar objects in Taurus (PI Schneider). The observations used the medium thickness optical blocking filter. XZ Tau and HL Tau were extracted using standard procedures in SAS version 19.1.0. Because of the close proximity of XZ Tau AB and HL Tau, we defined custom extraction regions to ensure minimal contamination of each source by the other source. The observations were energy filtered (0.3-8.0 keV) and time-filtered on good time intervals to reduce flaring particle background.

\subsection{AAVSO}

The two components of XZ Tau AB are known to be variable in the optical over the course of years, indicating potential outbursts expected of an ExOr object \citep{2008AJ....136.1980K}. To track the state of the XZ Tau AB and HL Tau systems, we requested observations of XZ Tau AB and HL Tau from the Association of Amateur Variable Star Observers (AAVSO) over the time periods of observation by Chandra and XMM. These observations, distributed across multiple observers, provide low-cadence optical monitoring over the course of the observations. While the AAVSO data do not resolve the separate components of XZ Tau A and B, \citet{2008AJ....136.1980K} resolves the two components with HST and find that XZ Tau A is relatively stable ($R$ magnitude changes $\sim 0.6$ mag between 1995 and 2004), while XZ Tau B can exhibit wide variations ($R$ magnitude changes $\sim 3.23$ mag between 2001 and 2004), which suggests that the bulk of the intrinsic variability is due to XZ Tau B.

%\begin{itemize}
%    \item Requested observations of XZ Tau and HL Tau from the AAVSO over the time period observed by Chandra and XMM
%    \item Provides low-cadence optical monitoring over the course of the time period of observation
%\end{itemize}

\section{Analysis}
\label{sec:analysis}

\subsection{Identifying Components}

We attempted to determine if the two components of XZ Tau AB were resolvable in the \textit{Chandra} data, using the positions for the stars recorded by ALMA \citep{2021ApJ...919...55I}. While XZ Tau A and B are separated by $\sim 0.3''$ \citep{2021ApJ...919...55I}, the \textit{Chandra} and \textit{XMM} point spread functions (PSFs) are too wide to precisely pinpoint whether the X-rays are coming from one star, or both--the center of the X-ray source in the \textit{Chandra} data is offset from both of the two ALMA sources ($0''.35$ north of XZ Tau A, $0''.26$ east of XZ Tau B). We thus treat all of our observations as the combined spectrum of both components.

% Position from srcdetect on reprocessed 21951:
% XZ Tau: 4:31:40.1049, +18:13:56.843

% From ALMA in 2017:
% XZ Tau A: 4:31:40.10, +18:13:56.63
% XZ Tau B: 4:31:40.08, +18:13:56.79

%\begin{itemize}
%    \item Positions from observations with ALMA
%    \item PSF in Chandra and XMM is wide enough that we cannot precisely pinpoint whether the X-rays are coming from one star, or both
%\end{itemize}

\subsection{Was XZ Tau AB in Outburst During the New X-Ray Observations?}

\begin{figure*}
    \centering
    \includegraphics[width=0.95\textwidth]{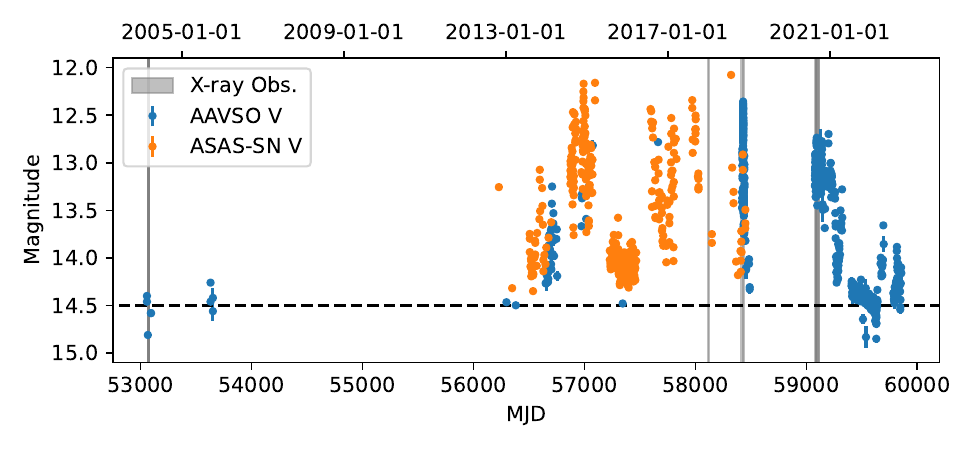}
    \caption{$V$-band light curve for XZ Tau AB from the AAVSO and ASAS-SN. The black dashed line represents the estimated baseline level for when XZ Tau B is not in an elevated state. Gray windows highlight times of X-ray observations in 2004, 2017, 2018, and 2020. ``Zoomed-in'' light curves for individual X-ray observations are shown in Figures \ref{fig:Chandra_lightcurves} and \ref{fig:XMM_lightcurves}. }
    \label{fig:full_optical_lightcurve}
\end{figure*}

We considered the full light curve for XZ Tau provided by the AAVSO, and used it to generate color light curves for each observation. Analyzing color as a function of time as well as the observed brightness should mitigate uncertainty in recorded brightness due to differences in observing setup for each observer. It also allows for evaluation of the extinction of the star as a function of time, which yields insight into the behavior of the circumstellar dust. To supplement the AAVSO light curves and serve as a ``ground truth'' measure of XZ Tau AB's brightness, we also considered the V-band light curve of XZ Tau from the ASAS-SN catalog of variable stars \citep{2014ApJ...788...48S,2019MNRAS.486.1907J}.

Based on \textit{Hubble Space Telescope} images of the resolved components of the XZ Tau system, \citet{2006A&A...453..241G} found that while XZ Tau B was potentially in outburst when the XZ Tau system was first observed with \textit{XMM-Newton} in 2000, it had clearly subsided by the time of the 2004 \textit{XMM-Newton} observations. \citet{2008AJ....136.1980K} found similar results.

The full AAVSO+ASAS-SN $V$-magnitude light curve for XZ Tau AB is presented in Figure \ref{fig:full_optical_lightcurve}, with the timeframes of various X-ray observations (including the 2018 \textit{Chandra} and 2020 \textit{XMM-Newton} observations) highlighted. We estimate the baseline ``quiescent'' level of the total $V$-band light from XZ Tau AB outside of outburst from the AAVSO data contemporaneous with the \textit{XMM-Newton} observations presented in \citet{2006A&A...453..241G}, as these observations were taken outside of outburst based on HST imaging that resolved both components of the system.

Over the twenty-year timeframe we consider, we see periods of quiescence of duration $\sim$ one year, in 2022. We also observe clear brightening events of durations that would suggest multiple EXOr-like outbursts, including ongoing outbursts during the 2018 \textit{Chandra} and 2020 \textit{XMM-Newton} observations. The 2017 \textit{Chandra} observations \citep{2020ApJ...888...15S} appear to take place during a local minimum of optical brightness, but at an elevated optical brightness compared to the baseline set in 2004.
%\pcs{I remember that we talked about the optical brightness, not the exact conclusions. However, I do have some doubts that the term "outburst" can be used if the source is during an "outburst" about 99\% of the time...?}

\subsection{X-ray Time Series}

The zeroth-order (non-dispersed) count rate for XZ Tau AB in the individual \textit{Chandra} observations is too low to produce a light curve with detailed time resolution. We instead look at the count rates for each observation as a whole.

\begin{figure}
    \centering
    \includegraphics[width=0.47\textwidth]{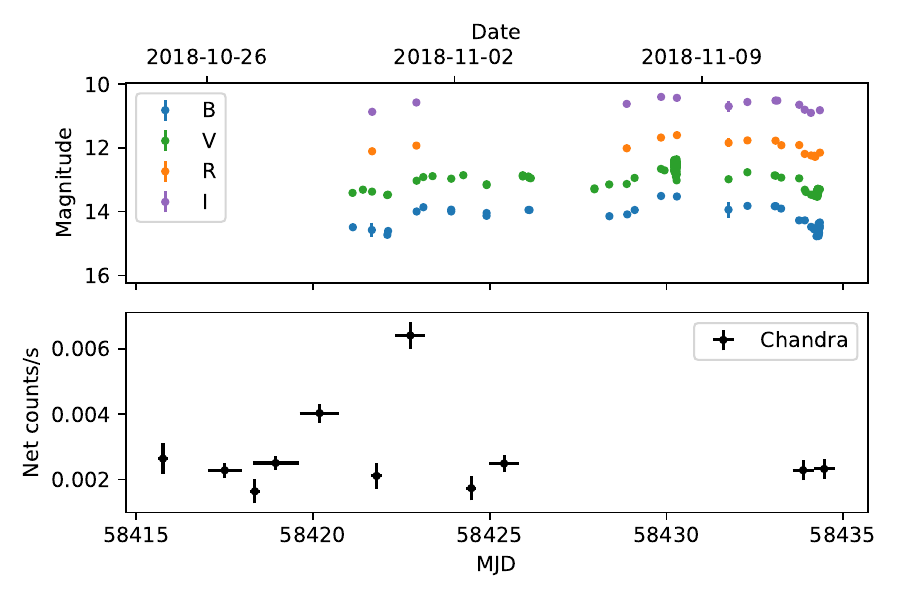}
    \caption{\emph{bottom:} Zeroth-order observation-averaged net count rates for each \textit{Chandra} observation of XZ Tau AB in Table \ref{tab:obs_sum}. Error bars in the time dimension indicate the duration of each observation. %The overall count rate is quite low due to contamination of the ACIS chip and the presence of the HETG.
    Two observations (obsids 20161 and 21951) show enhanced count rates compared to the other nine observations. \emph{top:} Contemporaneous AAVSO data shows some rolling variability with no obvious correlation to the \textit{Chandra} variability.}
    \label{fig:Chandra_lightcurves}
\end{figure}

We find that two of the eleven observations have significantly elevated count rates compared to the remaining nine, as depicted in Figure \ref{fig:Chandra_lightcurves}. For initial consideration, we treated these two bright observations separately, and combined the data from the remaining nine observations for improved statistics.

\begin{figure*}
    \centering
    \includegraphics[width=\textwidth]{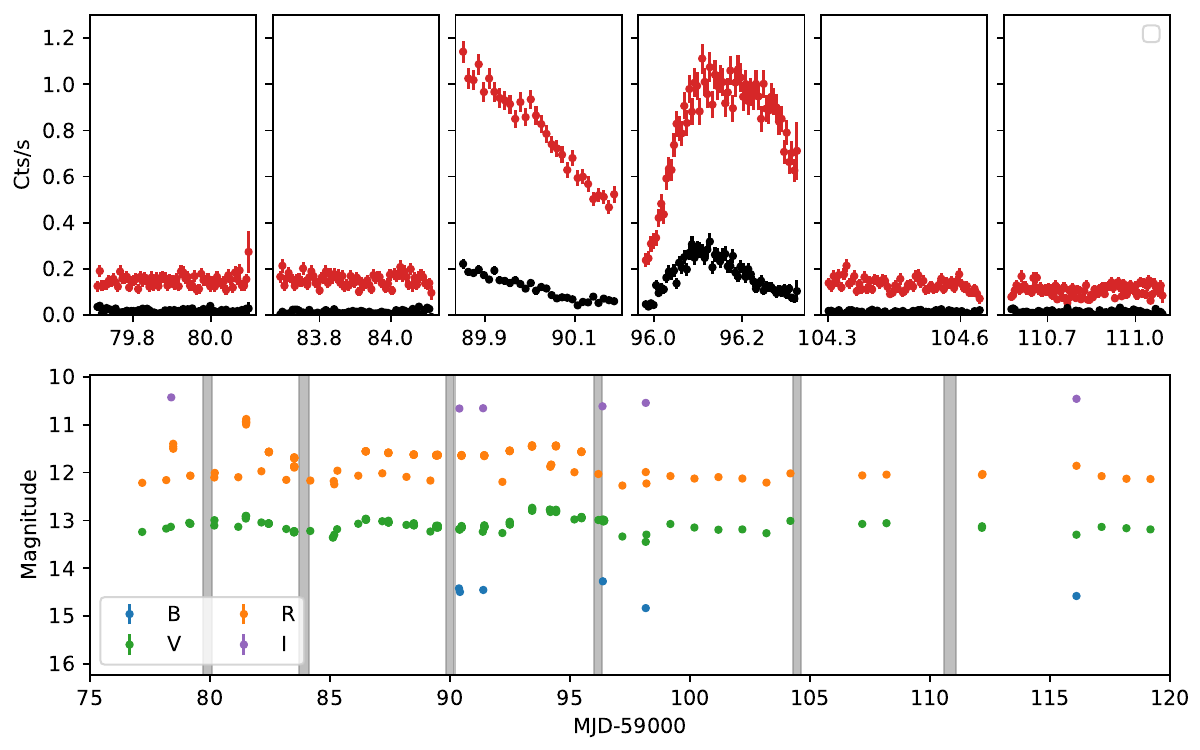}
    \caption{Broad-band light curves of XZ Tau AB from \textit{XMM-Newton} \emph{(top)} and AAVSO \emph{(bottom)}. The upper plots depict the individual \textit{XMM} observations in chronological order, presenting the hard- ($>1$ keV, black) and soft-energy ($<1$ keV, red) count rates from the EPIC-PN. The bottom light curve depicts observations of XZ Tau AB from AAVSO, with the \textit{XMM-Newton} observing windows highlighted in gray. Error bars are smaller than plot symbols. There is no apparent correlation in the X-ray and optical variability.}
    \label{fig:XMM_lightcurves}
\end{figure*}

We present light curves from \textit{XMM-Newton} with a 1000-s cadence in Figure \ref{fig:XMM_lightcurves}. The \textit{XMM-Newton} data have more counts than the \textit{Chandra} observations, enabling assessment of source variability on timescales shorter than the observation itself. Four of the \textit{XMM-Newton} observations are fairly stable in both low- and high-energy bands (defined as below and above 1 keV respectively). However, two observations are much brighter. The light curves of the six \textit{XMM-Newton} observations clearly indicate that the observations on 2020 August 28-29 and 2020 September 3-4 are flaring, with peak brightnesses $\sim 12\times$ the median for all observations in the hard band, and $\sim 5.5\times$ the median for all observations in the soft band, even after correcting for the high background at points during these observations. 

%\begin{itemize}
%    \item Comparison of AAVSO photometry (that does not resolve the binary) with past HST photometry (that does) indicates that the typically-fainter component is in a bright phase during these observations.
%    \item Looking at count rates from each obsid in Chandra shows two obsids that are significantly brighter than the other nine, in both high- and low-energy bands (defined as above and below 2 keV).
%    \item Similarly in XMM data there are two obsids that are significantly brighter in both high- and low-energy bands.
%    \item Count rate is too low with Chandra to assess evolution in detail
%    \item Count rate is high enough with XMM to identify that the bright obsids in XMM exhibit a flare-like temporal evolution within each obsid.
%\end{itemize}

\subsection{Modeling spectra}

We assume that the X-ray data are adequately described by two optically thin, collisionally excited emission components \citep[APEC models][]{2012ApJ...756..128F} in collisional equilibrium, assumed to be ``cool'' and ``hot,'' multiplied by a photoelectric absorption \texttt{phabs} component to model interstellar and circumstellar absorption. 

We also set neon and iron abundances as free parameters, and pin the abundances of nickel, silicon, calcium, and magnesium (which have similar first ionization potentials to iron) to the iron abundance. As a baseline, we fit the four ``quiescent'' observations from \textit{XMM-Newton} jointly, and use the temperatures, normalizations, absorption column density, and abundances from that fit as the initial guess for fitting each individual observation. We list the best-fit parameters for these models with free abundances in Table \ref{table:fitpars}.

Because of the diminished effective area of \textit{ACIS} at low energies caused by contamination, we fixed $N_{H}$ for the \textit{Chandra} observations at the best-fit $N_{H}$ for the quiescent \textit{XMM-Newton} observations to mitigate degeneracy between soft plasma and hydrogen absorption. We fit the zeroth-order spectrum for each ``bright'' \textit{Chandra} observation, but did not attempt to find individual fits for each of the faint \textit{Chandra} observations. Rather, we fit the combined zeroth-order spectrum for the other nine observations. We find that the model fits to the bright observations show no significant difference from the fit to the combined faint observations due to the substantial uncertainties that result from the limited number of counts. We therefore fit the combined zeroth-order spectrum from all eleven observations. In addition to fitting all spectra with free Ne and Fe abundances, we fixed Ne and Fe abundances at the best-fit values from the joint fit to the four quiescent \textit{XMM-Newton} observations and then fit all observations again; we list the best-fit parameters for these models with fixed abundances in Table \ref{table:fitparsFitted}. %To mitigate degeneracy between a soft plasma component and hydrogen absorption due to the diminished effective area of \textit{ACIS} at low energies caused by chip contamination, we fix the $N_{H}$ of the \textit{Chandra} data at the best-fit $N_{H}$ for the joint fit of the four faint \textit{XMM-Newton} observations. 

\begin{figure*}
    \centering
    \plottwo{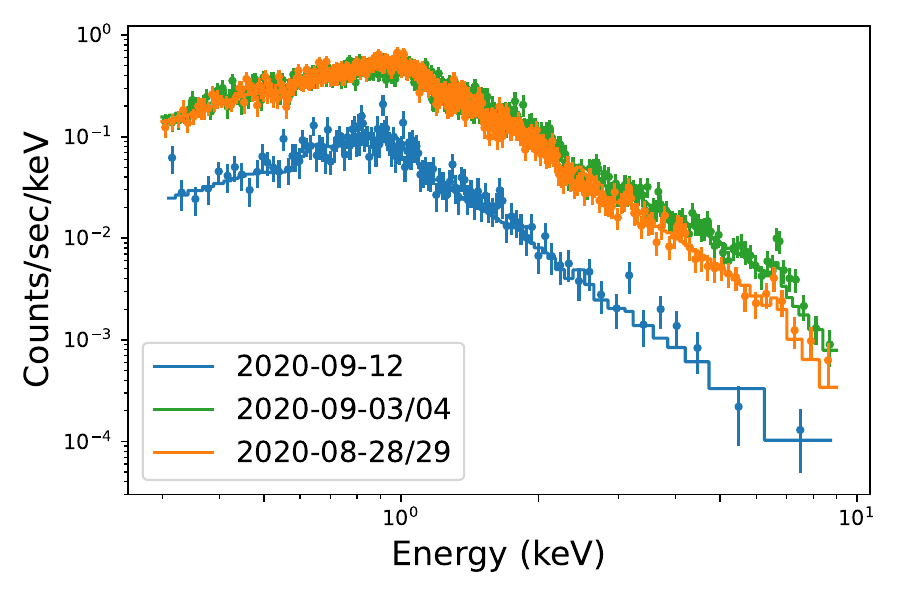}{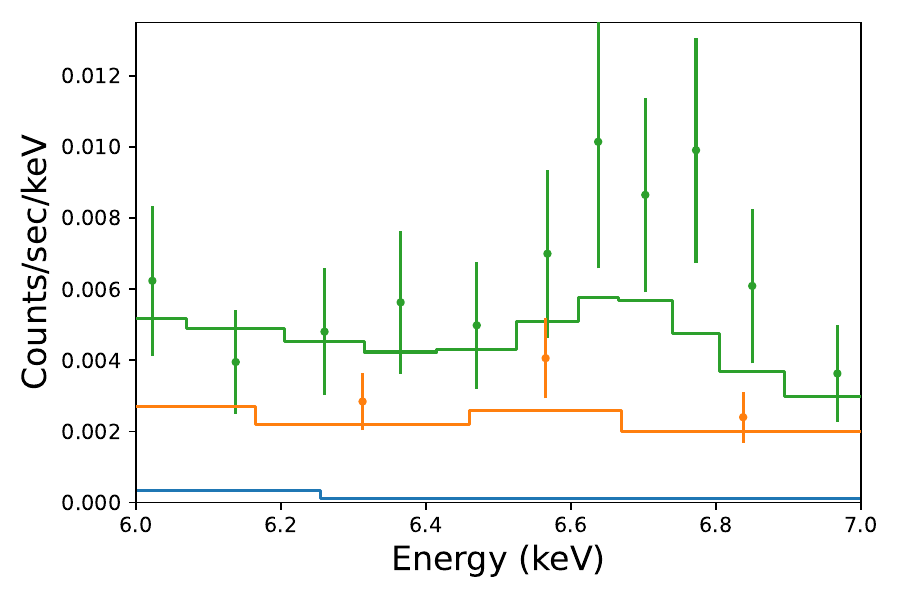}
    \caption{Spectrum of one quiescent observation with \textit{XMM-Newton} (blue), in comparison to the two observations with flares (orange and green). \textit{Left}: The two flare spectra exhibit enhancement in overall brightness, and show the Fe 6.7 keV emission line. \textit{Right}: A detailed look at the 6-7 keV range shows that the Fe 6.7 keV line exhibited by the flare spectra is underfit by models, and that flare spectra show no significant evidence for emission from the fluorescent Fe line at 6.4 keV.}
    \label{fig:XMM_spectrum_comparison}
\end{figure*}

We plot the PN spectra for one typical quiescent observation (obsid 0865040501) and the two flaring observations in Figure \ref{fig:XMM_spectrum_comparison}, with best fit models. As expected, the two flare spectra are brighter than the quiescent spectrum. Notably, the high-energy slopes of the flaring spectra are different from the quiescent spectrum, indicating a different high-energy state for those spectra. We explore time-resolved spectroscopy of the flares in Section \ref{sec:time_resolved_spectra}. This recalls a similar difference in spectral shape from 2000 to 2004 identified in \citet{2006A&A...453..241G}---a hard plasma component readily apparent in the \textit{XMM-Newton} spectrum from 2000 was no longer seen in 2004. Also notable in the flare observations is the presence of 6.7 keV emission from the Fe emission line. A closer look at the 6-7 keV range of the data (Figure \ref{fig:XMM_spectrum_comparison}, right) shows clear emission from the 6.7 keV line, but no significant evidence of emission from the fluorescent line at 6.4 keV, limited by the count rate and thus the bin widths adopted. It is notable that the two-temperature, fixed-abundance model here \textit{underpredicts} the Fe emission at 6.7 keV, albeit with marginal statistical significance. This can be explained by a change in Fe abundance from the quiescent to flaring observations, or by a lack of adequate high-energy emission in the simple 2-T model we adopt here.

\begin{deluxetable*}{lccccccccc}
\tablecaption{Summary of parameters for best fit models for X-ray spectra of XZ Tau AB with fixed Ne and Fe \label{table:fitparsFitted}}
\tablehead{ & & \colhead{$N_{H}$} & \multicolumn{2}{c}{kT (keV)} & \multicolumn{2}{c}{EM ($10^{52}\, \mathrm{{cm}^{-3}}$)} & \multicolumn{2}{c}{Flux ($10^{-13} \mathrm{erg\,cm^{-2}\,s^{-1}}$)} & \colhead{Log($L_{X}$)}\\
\colhead{Source} & \colhead{rstat\tablenotemark{a}} & \colhead{($10^{22} \mathrm{cm^{-2}}$)} & \colhead{Cool}  & \colhead{Hot}  & \colhead{Cool}  & \colhead{Hot}  & \colhead{Absorbed} & \colhead{Unabsorbed} & \colhead{ ($\mathrm{erg\,s^{-1}}$)}}
\startdata
Faint \textit{Chandra}\tablenotemark{b} & 1.025  & $0.113                  $ & $0.48_{-0.08}^{+0.07}$ & $2.0_{-0.3}^{+ 0.3}$ & $ 3.5_{-0.7}^{+0.6}$ & $ 3.0_{-0.5}^{+0.6}$ &   2.14 &  2.93 & 29.8 \\
201610     & 0.3565 & $0.113                  $ & $0.7 _{-0.5 }^{+0.2 }$ & $> 2.2             $ & $ 9  _{-4  }^{+2  }$ & $ 2.4_{-0.9}^{+0.6}$ &   4.12 &  5.44 & 30.1 \\
219510     & 0.1261 & $0.113                  $ & $0.7 _{-0.4 }^{+0.4 }$ & $3  _{-3  }^{+65  }$ & $13  _{-6  }^{+4  }$ & $ 5  _{-4  }^{+7  }$ &   6.06 &  8.12 & 30.3 \\
All \textit{Chandra}  & 0.7606 & $0.113                  $ & $0.49_{-0.05}^{+0.10}$ & $2.3_{-0.3}^{+ 0.4}$ & $ 4.6_{-0.6}^{+0.7}$ & $ 3.6_{-0.7}^{+0.5}$ &   2.84 &  3.86 & 30.0 \\
Faint \textit{XMM-Newton}\tablenotemark{c}   & 0.9135 & $0.113_{-0.007}^{+0.007}$ & $0.75_{-0.02}^{+0.02}$ & $3.2_{-0.3}^{+ 0.5}$ & $ 4.3_{-0.2}^{+0.2}$ & $ 1.7_{-0.2}^{+0.2}$ &   2.08 &  2.74 & 29.8 \\
0865040201 & 0.6056 & $0.09 _{-0.01 }^{+0.01 }$ & $0.72_{-0.04}^{+0.04}$ & $4.6_{-0.9}^{+ 1.6}$ & $ 4.3_{-0.3}^{+0.3}$ & $ 2.0_{-0.3}^{+0.3}$ &   2.48 &  3.06 & 29.9 \\
0865040301 & 0.5751 & $0.077_{-0.009}^{+0.009}$ & $0.84_{-0.03}^{+0.03}$ & $5.4_{-2.3}^{+14.4}$ & $ 5.4_{-0.4}^{+0.3}$ & $ 0.9_{-0.2}^{+0.3}$ &   2.33 &  2.83 & 29.8 \\
0865040401 & 0.7459 & $0.127_{-0.005}^{+0.005}$ & $0.87_{-0.02}^{+0.02}$ & $2.8_{-0.1}^{+ 0.2}$ & $21  _{-1  }^{+1  }$ & $23  _{-1  }^{+1  }$ &  16.3  & 20.8  & 30.7 \\
0865040601 & 0.7041 & $0.105_{-0.005}^{+0.005}$ & $0.85_{-0.03}^{+0.03}$ & $3.8_{-0.2}^{+ 0.2}$ & $13.4_{-0.8}^{+0.8}$ & $31  _{-1  }^{+1  }$ &  19.8  & 23.3  & 30.7 \\
0865040701 & 0.6863 & $0.10 _{-0.01 }^{+0.01 }$ & $0.77_{-0.04}^{+0.04}$ & $2.9_{-0.5}^{+ 0.8}$ & $ 4.0_{-0.4}^{+0.4}$ & $ 2.0_{-0.4}^{+0.3}$ &   2.15 &  2.75 & 29.8 \\
0865040501 & 0.6152 & $0.15 _{-0.01 }^{+0.02 }$ & $0.71_{-0.04}^{+0.04}$ & $2.9_{-0.5}^{+ 0.7}$ & $ 3.9_{-0.3}^{+0.3}$ & $ 1.6_{-0.3}^{+0.3}$ &   1.72 &  2.50 & 29.8 \\
\enddata
\tablenotetext{a}{Reduced $\chi^{2}$ using Gehrels weighting.}
\tablenotetext{b}{Joint fit of obsids 20160, 21946, 21947, 21948, 21950, 21952, 21953, 21954, and 21965.}
\tablenotetext{c}{Joint fit of obsids 0865040201, 0865040301, 0865040501, and 0865040701.}
\end{deluxetable*}

\subsection{Time-resolved Spectra of X-ray Flares}
\label{sec:time_resolved_spectra}

\citet{2007A&A...468..485F} present a time-resolved spectroscopic analysis of an observation of XZ Tau AB originally presented in \citet{2003A&A...403..187F} and reanalyzed in \citet{2006A&A...453..241G}. In this observation, XZ Tau AB monotonically rises over the course of the $\sim55$ ks observation. By contrast, in obsid 0865040401 (hereafter referred to as 401), XZ Tau AB monotonically decays, and in observation 0865040601 (hereafter referred to as 601), it exhibits a sharp rise and then decay. The short timescales involved indicate that these are typical flares, rather than long time-scale increases in X-ray brightness associated with magnetic reconnection flares in FUor/EXor outbursts \citep{2004Natur.430..429K}.

\begin{figure}
    \centering
    \includegraphics[width=0.47\textwidth]{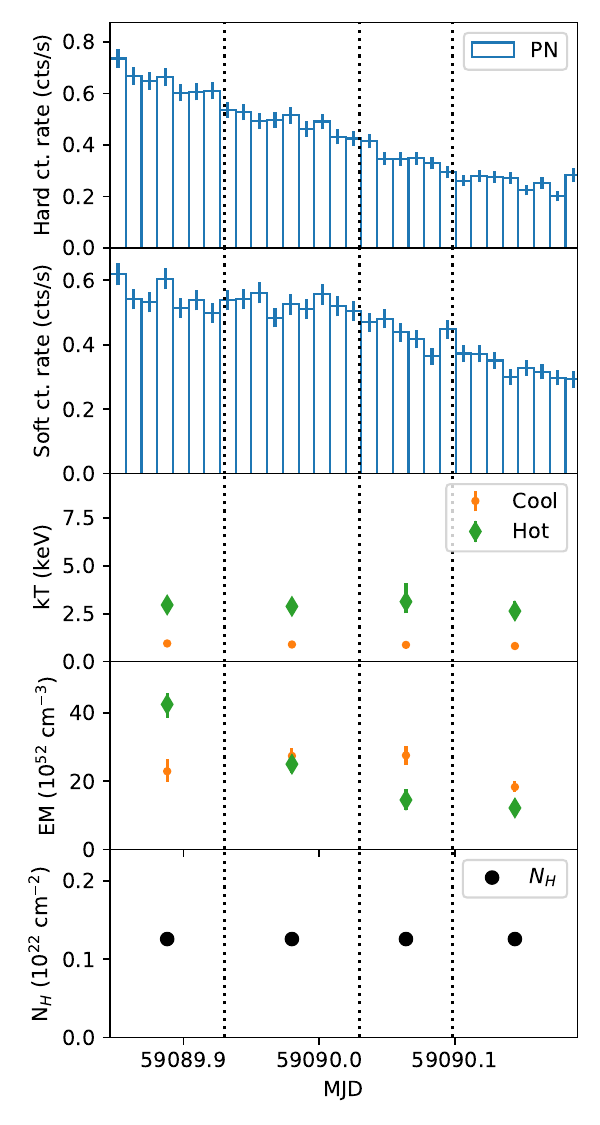}
    \caption{Evolution of XZ Tau AB over the course of a flare decay in observation 0865040401. The decrease in the emission measure of the ``hot'' component over time clearly tracks the decrease in count rate for photons at energies $>1$ keV, while the temperature of the ``hot'' component remains stable.}
    \label{fig:lightcurve_with_params_401}
\end{figure}

\begin{figure}
    \centering
    \includegraphics[width=0.47\textwidth]{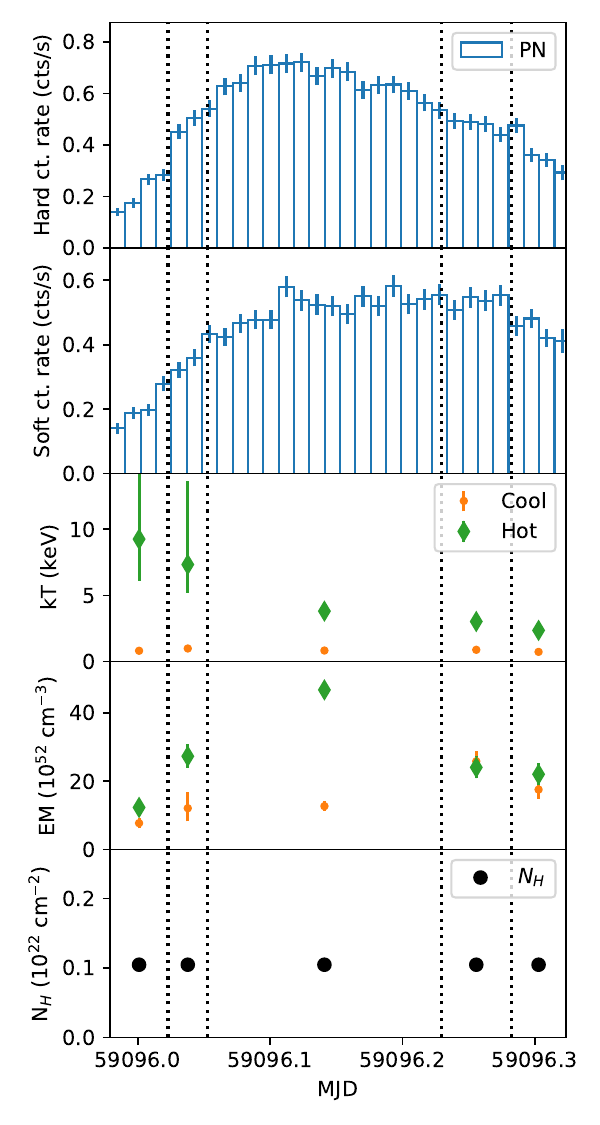}
    \caption{Evolution of XZ Tau AB over the course of a flare rise and decay in observation 0865040601. The change in the emission measure of the ``hot'' component over time clearly tracks the rise in count rate for photons at energies $>1$ keV. The ``cool'' component stays fairly stable over time, indicating that most of the change in brightness is due to the hot component.}
    \label{fig:lightcurve_with_params_601}
\end{figure}

Following the technique of \citet{2007A&A...468..485F}, we divide observations 401 and 601 each into blocks of time using the \texttt{bayesian\_blocks} method as provided in AstroPy, and fit models to the spectra from each of these blocks of time to better understand the flare evolution. We assume in these fits that (a) the Ne and Fe-like abundances match those jointly fit to the quiescent observations, and that (b) the hydrogen column density does not change over the course of one observation (i.e.\ it remains fixed at the best-fit value from fitting the spectrum of the full-duration observation). We present the best-fit parameters in Table \ref{table:fitparsSplinter}, and plot the evolution in parameters as a function of time along with the observation light curves in Figures \ref{fig:lightcurve_with_params_401} and \ref{fig:lightcurve_with_params_601}. We discuss the evolution of the flares observed in these observations in more detail in Section \ref{sec:flares}. 

\subsection{Gratings Data}

\begin{figure*}
    \centering
    \includegraphics[width=\textwidth]{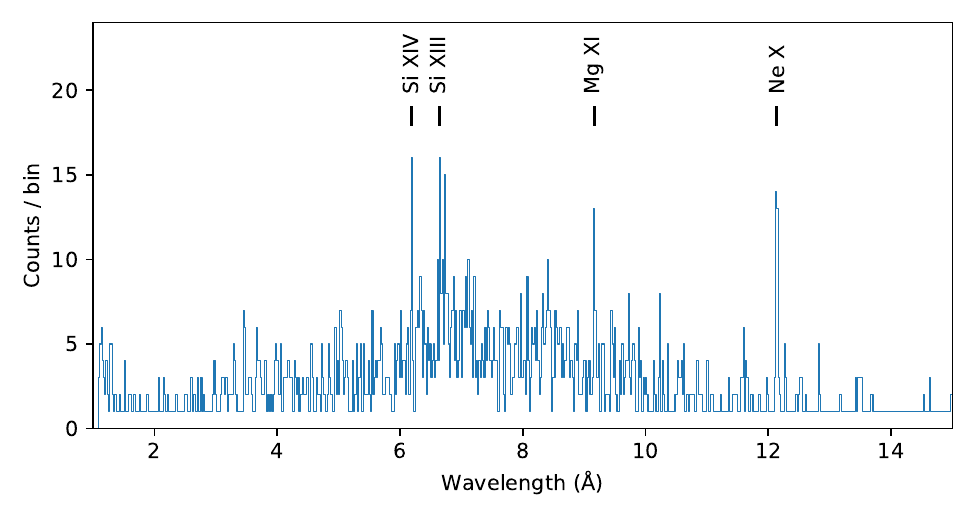}
    \caption{\textit{Chandra}/HETG data for XZ Tau AB, combining both the MEG and HEG. Prominent lines in the spectrum are labeled. We exclude the spectrum at wavelengths longer than 15 \AA as there is minimal emission at these wavelengths due to the contamination of the ACIS-S chip.}
    \label{fig:gratings}
\end{figure*}

The combined first (positive and negative) order gratings data from both the High-Energy and Medium-Energy Gratings (HEG and MEG, respectively) from the eleven \textit{Chandra} observations are presented in Figure \ref{fig:gratings}. We do not consider gratings data from \textit{XMM-Newton}, due to the likely blend of data in the RGS from XZ Tau AB and the nearby (separation $\sim 12''$) HL Tau.  Following the method presented in \citet{2021ApJ...915..114P}, we fit the \textit{Chandra} grating spectra with a two-temperature model as obtained from \textit{XMM-Newton} fits, while allowing the spectral parameters to vary over the range allowed by \textit{XMM-Newton} and fixing the abundances to the \textit{XMM-Newton} values. We then fit the individual regions line-by-line, grouping the Si XIII, Mg XII, Mg XI, Ne X, Ne IX by a factor of 2. We list our line fluxes for the gratings data in Table \ref{tab:linefluxes}. Due to the low signal in the Mg XI and Ne IX triplet features, lines typically used to differentiate low-density coronal plasma from high-density plasma associated with accretion shocks, we are unable to constrain the plasma density.

%\begin{itemize}
    %\item Setting abundances based on the XMM spectra, fitting accordingly.
    %\item Distinct differences in abundance for the `flaring' spectra
    %\item Tables where we show how the characteristics of each obsid differ depending on assumptions (freely evolving Ne/Fe; fixed Ne/Fe at the best-fit abundances from faint-XMM with free nH, temperatures, norms; fixed Ne/Fe and nH from the faint-XMM obsids with free temperatures and norms; fixed Ne/Fe, nH, and temperatures from faint-XMM with varying norms)
    %\item Plots showing evolution over time with associated commentary
 %   \item Line ratios from Chandra to better constrain the characteristics of the plasma (as best we can, anyway).
%\end{itemize}

\begin{deluxetable}{lcc}
\tablecaption{Line Fluxes from HETGS Spectrum of XZ Tau \label{tab:linefluxes}}
\tablehead{\colhead{} & \colhead{Wavelength} & \colhead{Flux} \\ 
\colhead{Line} & \colhead{(\AA)} & \colhead{($10^{-7}$ photons cm$^{-2}$ s$^{-1}$)}}
\startdata
Si XIII (r) &  6.648 & $ 3^{ +3}_{ -2}$ \\
Si XIII (i) &  6.687 & $<3.36$ \\
Si XIII (f) &  6.740 & $ 3^{ +2}_{ -2}$ \\
Mg XII      &  8.422 & $<3.63$ \\
Mg XI   (r) &  9.169 & $ 9^{ +5}_{ -4}$ \\
Mg XI   (i) &  9.230 & $ <2.13 $ \\
Mg XI   (f) &  9.314 & $ 3^{ +4}_{ -3}$ \\
Ne X        & 12.135 & $50^{+21}_{-17}$ \\
Ne IX   (r) & 13.447 & $<29.7$ \\
Fe XIX      & 13.462 & $28^{+33}_{-22}$ \\
Fe XIX      & 13.518 & $14^{+24}_{- 7}$ \\
Ne IX   (i) & 13.552 & $<18.1$ \\
Fe XIX      & 13.645 & $<27.7$ \\
Ne IX   (f) & 13.699 & $17^{+24}_{-13}$ \\
Fe XX       & 13.767 & $<1.32$ \\
Fe XVII     & 13.825 & $<2.16$ \\
\enddata
\end{deluxetable}

\section{Discussion}
\label{sec:discussion}

\subsection{X-ray Evolution Over Time, Or Lack Thereof}
\label{sec:stable}

%\begin{figure}
%    \centering
%    \includegraphics[width=0.48\textwidth]{figures/kT_vs_time.pdf}
%    \caption{Temperature of each component as a function of time for each observation. Data from \textit{Chandra} is presented in the left panel, while data from \textit{XMM-Newton}, collected two years later, is presented in the right panel. The cool component remains remarkably stable over time, while the hot component only significantly varies during the flares.}
%    \label{fig:kT_vs_time}
%\end{figure}

%\begin{figure}
%    \centering
%    \includegraphics[width=0.48\textwidth]{figures/EM_vs_time.pdf}
%    \caption{Emission measure for each component as a function of time for each observation. Data from \textit{Chandra} is presented in the left panel, while data from \textit{XMM-Newton}, collected two years later, is presented in the right panel. The cool component remains remarkably stable over time, while the hot component only significantly varies during the flares.}
%    \label{fig:EM_vs_time}
%\end{figure}

We present the best-fit temperatures, emission measures, and abundances for each observation in Table \ref{table:fitpars}. We note that within uncertainties the two components exhibit remarkable consistency in temperature over time. The cool component in particular shows little variation in temperature. The hot component exhibits significantly more uncertainty in each observation, such that while the best-fit temperatures can differ by $\sim 1 \mathrm{\;keV}$, the temperatures are generally consistent with each other within the uncertainties. The significant variation in brightness during flares appears to come solely from the emission measure itself, which exhibits moderate increase in the cool component during a flare and an increase on order $\sim30$ in the hot component. This is generally consistent with a model of occasional flares superposed on top of a stable underlying stellar corona.

\begin{figure*}
    \centering
    \includegraphics[width=\textwidth]{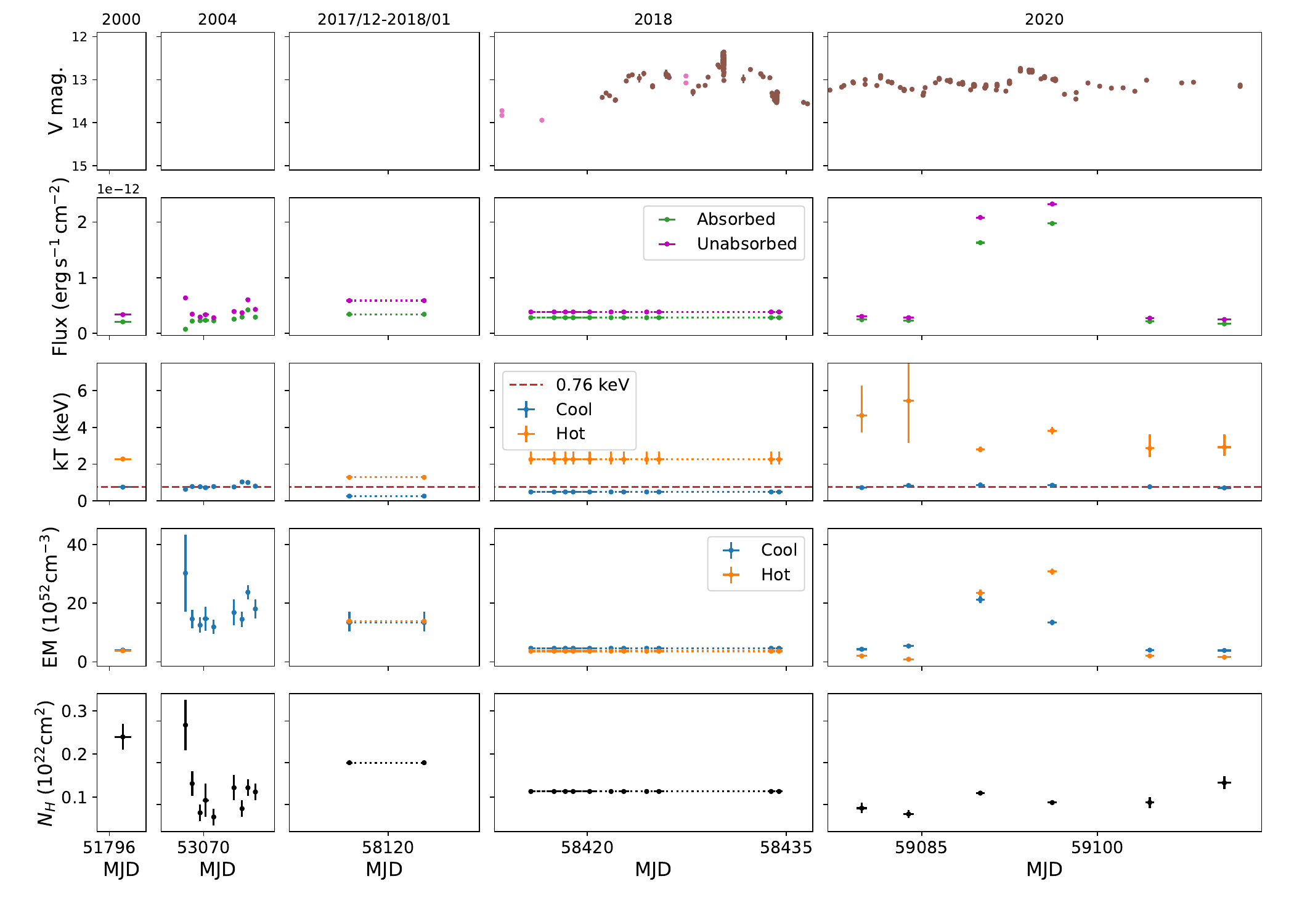}
    \caption{Characteristics of best-fit models to various observations of XZ Tau AB over time, from published work and analysis of unpublished publicly-available data. Rows display (top to bottom) optical brightness as represented by $V$ magnitude, X-ray flux (absorbed and unabsorbed), plasma temperatures, plasma emission measures, and $N_{H}$. Each column spans one set of observations, from 2000 \citep{2007AandA...468..353G}, 2004 \citep{2006A&A...453..241G}, 2017-2018 \citep{2020ApJ...888...15S}, 2018, and 2020. Dotted lines connect points representing obsids that were jointly fit.}
    \label{fig:previous_work_vs_time}
\end{figure*}

To consider the hypothesis of a stable underlying stellar corona, we compare our models to those of previous work analyzing X-ray spectra from XZ Tau AB. We present our results and previous work as a function of time in Figure \ref{fig:previous_work_vs_time}, and briefly discuss the relevant parameters below. The parameters from previous work are summarized in the Appendix.

\citet{2020ApJ...888...15S} jointly fit their two observations of XZ Tau AB in 2017 December and 2018 January with a two-component model with a fixed hydrogen column density of $N_{H} = 0.2 \mathrm{cm^2}$. They find best-fit model temperatures that are quite cool (0.23 keV and 1.28 keV, respectively) compared to those we find. \citet{2020ApJ...888...15S} also present a norm-weighted average temperature of these two components of 0.78 keV, consistent with our cool temperature, and the normalizations of those two components are similarly consistent. This indicates to us that the two-component model they adopt is essentially reproducing our cool component. Similarly, with carefully chosen initial parameters for fitting our data, we can produce a fit to our faint \textit{XMM-Newton} data with similar (though somewhat worse) goodness-of-fit with temperatures similar to the temperatures found in this work. We thus conclude that the two temperatures found by \citet{2020ApJ...888...15S} reproduce our cool component.

\citet{2007AandA...468..353G} present multiple fit options to all sources in the \textit{XMM-Newton} Extended Survey of the Taurus Molecular Cloud (XEST), including a two-temperature plasma model similar to the parameterization we adopt. They present a fit to the ``characteristic'' non-flare interval in the data \citep[also analyzed in][]{2003A&A...403..187F,2006A&A...453..241G}, with best-fit temperatures of $\mathrm{kT_{1}} = 0.75$ keV and $\mathrm{kT_{2}} = 2.28$ keV, respectively, with emission measures $\mathrm{EM_{1}} = 4.00 \times 10^{52} \mathrm{cm}^{-3}$ and $\mathrm{EM_{2}} = 3.76 \times 10^{52} \mathrm{cm}^{-3}$, and a hydrogen column density of $N_{H} = 0.24 \pm 0.03 \times 10^{22} \mathrm{cm^{2}}$. \citet{2007A&A...468..485F} split the observation used for XZ Tau AB in \citet{2007AandA...468..353G} into five time-resolved spectra using the Bayesian blocks method and fit a two-temperature plasma to each of these. In this paradigm, the brightening of the light curve corresponded to an increase in the emission measure of the hot plasma, while the cool component held at constant temperature and (until the last spectrum) emission measure. These data directly correspond to our fits despite differences in the treatment of abundance. While \citet{2003A&A...403..187F} identify a significant decrease in emission measure of the cool component for this observation, \citet{2006A&A...453..241G} note that this is due to the spurious elevated $N_\mathrm{H}$ level in the first time bin of the time-resolved analysis. Table 4 in \citet{2006A&A...453..241G} \citep[which supersedes the Table in ][]{2003A&A...403..187F} shows that while the emission measure does increase, this increase is not statistically significant. We find that our variability is consistent with the variability summary in \citet{2003A&A...403..187F} as well; our observed fluxes fall neatly in the range of observed fluxes in this older data. 

\citet{2006A&A...453..241G} present five days of monitoring of a field that includes XZ Tau AB with \textit{XMM-Newton}, during a period of high background for the telescope. They present single-temperature fits to these data, finding a stable plasma (albeit with changing hydrogen column density) at a temperature between 0.63 keV and 0.8 keV. Fits to two of their observations indicate a hotter temperature; however, these also show an increase in the observed flux, suggesting a flare. Indeed, the authors note that some of these observations are better fit by a two-component model. We argue that this too demonstrates the persistence of a cool component between 0.7 keV and 0.8 keV over time. We also note that while \citet{2006A&A...453..241G} hypothesize that the change in the XZ Tau AB X-ray spectrum is connected to the end of XZ Tau B's outburst, we find that the spectrum remains similar to the 2004 result despite the apparent outburst of XZ Tau B during our observations.

From these data, we see that both in the raw observables and in the adopted modeling paradigm of stellar X-ray spectra as stacked single-temperature plasmas, XZ Tau AB remains fairly stable over observations spanning the 2000-2020 timeframe, with the few exceptions of likely flares, as shown in Figure \ref{fig:previous_work_vs_time}. Observed fluxes remain fairly stable over time apart from the observations we identify as flares. $N_{H}$ varies around a consistently low level, indicating that the observations (particularly recent \textit{Chandra} observations with limited sensitivity to soft X-rays) do not have much leverage to constrain $N_H$. The best-fit models consistently exhibit a cool temperature between 0.7-0.8 keV, indicating a stable temperature feature we interpret as the stellar corona(e). Due to different models assumptions (e.g.\ abundances), the emission measures and $N_\mathrm{H}$ vary widely for different datasets and are not directly comparable. The ratio of temperature components and variability \textit{within} each set are robust, as can be seen in the flares in 2020. The consistency of the cool component over an interval of \textit{twenty years} bolsters our hypothesis of an underlying stable corona, with occasional flares superposed on top of this, and recommends against the idea that flares completely reorganize the stellar coronal behavior.

\begin{deluxetable*}{lccccccccc}
\tablecaption{Summary of parameters for best fit models for time-resolved spectra of XZ Tau AB Flares \label{table:fitparsSplinter}}
\tablehead{\colhead{} & \colhead{} & \colhead{$N_{H}$} & \multicolumn{2}{c}{Cool Component} & \multicolumn{2}{c}{Hot Component} & \multicolumn{2}{c}{Flux ($10^{-13} \mathrm{erg\,s^{-1}\,cm^{-2}}$)} & \colhead{Log(Lum.)} \\
\colhead{obsid} & \colhead{rstat} & \colhead{($10^{22} \mathrm{cm}^{-2}$)} & \colhead{kT (keV)} & \colhead{EM ($10^{52} \mathrm{cm}^{-3}$)} & \colhead{kT (keV)} & \colhead{EM ($10^{52} \mathrm{cm}^{-3}$)} & \colhead{Absorbed} & \colhead{Unabsorbed} & \colhead{log (erg $\mathrm{s}^{-1}$)}}
\startdata
FaintXMM    & 0.9135 & $ 0.113 _{ -0.007 }^{ +0.007 } $ & $ 0.75 _{ -0.02 }^{ +0.02 } $ & $  4.3 _{ -0.2 }^{ +0.2 } $ & $ 3.2   _{ -0.3 }^{ +0.5 } $ & $  1.7 _{ -0.2 }^{ +0.2 } $ &  2.078 &  2.739 & 29.8 \\
0865040401A & 0.5724 & $ 0.127 $ & $ 0.95 _{ -0.05 }^{ +0.06 } $ & $ 21   _{ -3   }^{ +3   } $ & $ 3.0   _{ -0.2 }^{ +0.3 } $ & $ 39   _{ -3   }^{ +3   } $ & 23.98  & 28.71  & 30.8 \\
0865040401B & 0.7251 & $ 0.127 $ & $ 0.89 _{ -0.04 }^{ +0.04 } $ & $ 25   _{ -2   }^{ +2   } $ & $ 2.9   _{ -0.3 }^{ +0.4 } $ & $ 23   _{ -2   }^{ +2   } $ & 17.92  & 22.08  & 30.7 \\
0865040401C & 0.5207 & $ 0.127 $ & $ 0.87 _{ -0.04 }^{ +0.04 } $ & $ 25   _{ -3   }^{ +2   } $ & $ 3.1   _{ -0.6 }^{ +1.0 } $ & $ 13   _{ -3   }^{ +3   } $ & 13.92  & 17.49  & 30.6 \\
0865040401D & 0.6073 & $ 0.127 $ & $ 0.81 _{ -0.03 }^{ +0.04 } $ & $ 17   _{ -1   }^{ +2   } $ & $ 2.6   _{ -0.4 }^{ +0.5 } $ & $ 11   _{ -2   }^{ +2   } $ &  9.909 & 12.54  & 30.5 \\
0865040601A & 0.5601 & $ 0.105 $ & $ 0.82 _{ -0.12 }^{ +0.10 } $ & $  7   _{ -1   }^{ +1   } $ & $ 9     _{ -3   }^{ +9   } $ & $ 11   _{ -1   }^{ +1   } $ &  9.291 & 10.73  & 30.4 \\
0865040601B & 0.6247 & $ 0.105 $ & $ 0.99 _{ -0.15 }^{ +0.14 } $ & $ 11   _{ -3   }^{ +4   } $ & $ 7     _{ -2   }^{ +6   } $ & $ 25   _{ -3   }^{ +3   } $ & 18.65  & 21.23  & 30.7 \\
0865040601C & 0.6304 & $ 0.105 $ & $ 0.84 _{ -0.05 }^{ +0.05 } $ & $ 12   _{ -1   }^{ +1   } $ & $ 3.8   _{ -0.2 }^{ +0.2 } $ & $ 42   _{ -1   }^{ +1   } $ & 25.03  & 29.13  & 30.8 \\
0865040601D & 0.7071 & $ 0.105 $ & $ 0.89 _{ -0.05 }^{ +0.05 } $ & $ 23   _{ -3   }^{ +3   } $ & $ 3.0   _{ -0.4 }^{ +0.6 } $ & $ 22   _{ -3   }^{ +3   } $ & 17.32  & 21.27  & 30.7 \\
0865040601E & 0.591  & $ 0.105 $ & $ 0.74 _{ -0.07 }^{ +0.06 } $ & $ 16   _{ -2   }^{ +2   } $ & $ 2.4   _{ -0.4 }^{ +0.5 } $ & $ 20   _{ -3   }^{ +3   } $ & 13.07  & 16.38  & 30.6
\enddata
\end{deluxetable*}

\subsection{Short-term Coronal Variability}
\label{sec:flares}

Observation 401 shows the clear signatures of the decay phase of a flare. The hardness of the spectrum evolves over time--the count rates clearly show a decrease in hard flux while soft flux remains stable over the first two blocks of time (before roughly MJD 59090.03), before the soft flux also begins to decay over the latter half of the observation. The best-fit temperatures, abundances, and hydrogen column densities remain stable across the observation, though there is substantial uncertainty in the temperature of the hot component in the third, shortest, block of time. The clearest change over time is in the emission measures of the ``hot'' component, which we will refer to in this section as the flare component. The flare component's emission measure decreases by $>60\%$ between the first and fourth blocks, while the emission measure of the cool component stays stable (within uncertainties) over that time frame. This indicates that for this flare, the decay is solely driven by the emission measure decreasing. It is notable that the emission measures invert in the third block of time, albeit with substantial uncertainties.

Observation 601, on the other hand, shows both the rise and the start of the decay for the flare, and the characteristics are not nearly as clear-cut as in 401. Qualitatively, as the hard flux peaks the soft flux levels off, leading to a decay phase similar to that seen in 401. The increase in hard flux is steeper than in the soft, as there is less hard flux initially and more hard flux at the peak.

While the cool component maintains its characteristic behavior (albeit with substantial uncertainties) throughout the flare, the hot component is far more variable. During the rise phase, the temperature is elevated compared to both 401 and the decay phase of this flare, while the emission measure increases, suggesting a hotter plasma in the rise than in the decay.

\begin{figure}
    \centering
    \includegraphics[width=0.47\textwidth]{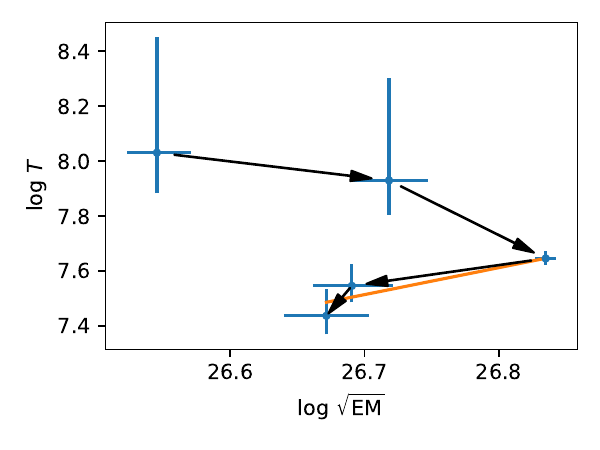}
    \caption{$\log(T)$ of the five time-resolved spectra of  Observation 601, as a function of $\log(\sqrt{EM})$. Arrows indicate the time order of observations. The orange line reflects the slope $\zeta$ of the flare decay in this space.}
    \label{fig:T_vs_EM}
\end{figure}

The detection of the flare peak in observation 601 allows for a fuller characterization of the coronal loops produced by the flare based on the flare decay. Following the application of \citet{2004A&A...416..733R} by \citet{2006A&A...453..241G} and \citet{2007A&A...468..485F}, we estimate the flare loop semi-length $L$ from the $e$-folding timescale of the decay and the maximum temperatures $T_{\mathrm{max}}$ via the equation

\begin{equation}
L = \frac{\tau_{\mathrm{LC}} \sqrt{T_{\mathrm{max}}}}{\alpha F(\zeta)},
\label{eq:L}
\end{equation}

%3.7 × 10−4 cm−1 s−1 K 1/2 

where $\alpha = 3.7 \times 10^{-4} \mathrm{cm^{-1}\,s^{-1}\,K^{1/2}}$, $\tau_{\mathrm{LC}}$ is the $e$-folding timescale of the decay, and $T_{\mathrm{max}}$ is the maximum plasma temperature of the flare, while accounting for heating during the decay via $\zeta$, the slope of the flare decay in the $\log(T)$ vs. $\log(\sqrt{EM})$ plane, shown in Figure \ref{fig:T_vs_EM}.

These characteristics (including the relationship between the best-fit peak temperature $T_{\mathrm{obs}}$ and $T_{\mathrm{max}}$) are calibrated for the PN detector; however, \citet{2007A&A...468..485F} found that these formulae would give order-of-magnitude results for the MOS detectors as well, given the similarities of the instruments and the width of the adopted spectral band. We thus use these equations for our temperatures and emission measures derived from joint fits to the three detectors.

For the flare in 601, we find $\zeta = 1.2 \pm 0.4$. We find an $e$-folding timescale based on fitting a line to the natural logarithm of the decay phase light curve of $50 \pm 4$ ks. The best-fit maximum temperature $T_{\mathrm{obs}} = 44_{-2}^{+3}$ MK, which following Equation (3) from \citet{2006A&A...453..241G} corresponds to a maximum temperature of $T_\mathrm{max} = 96^{+4}_{-3}$ MK. Inputting this information into Equation \ref{eq:L} yields a loop length of $\sim 10 R_{\odot}$, corresponding to $\sim 6 R_{\star}$ for XZ Tau A and $\sim 9 R_\star$ for XZ Tau B. These lengths are such that the flare \textit{could} reach into the disk, though we do not have direct measurement of the inner disk radius for either star; ALMA data do not resolve the dust emission for either disk and thus leave the radius as $\lesssim 15$ au \citep{2021ApJ...919...55I}. Such extended flare sizes are not uncommon in pre-main sequence disk-hosting stars \citep[e.g.][]{2005ApJS..160..469F}; however, there is no apparent difference in flare energy or flare peak energy between flares on disk-hosting and diskless pre-main sequence stars \citep{2021ApJ...916...32G}, and further work suggests that flares in disk-hosting pre-main sequence stars also exhibit loops with both footprints in the stellar surface, rather than a flare that extends from star to disk \citep{2021ApJ...920..154G}. The data in hand do not make it more or less likely that either star produced the flare, in our opinion; while XZ Tau B is thought to be an ExOR object, both XZ Tau A and B are young stars with disks, making flare events likely from both.

We also attempted using the available data from the flare during observation 401 to provide a lower limit for the coronal loops in that flare as well. For that flare, we find a lower limit decay $e$-folding time of $33 \pm 1$ ks and a lower limit maximum temperature of $72_{-4}^{+5}$ MK. However, for this flare $\zeta \sim$ 0.1, well below the validity threshold for the relation of $0.35<\zeta\leq1.6$.

\subsection{A YSO Coronal Spectrum Over Time Looks like a Snapshot of a Young Cluster}

Many studies of young stars, such as the Chandra Orion Ultradeep Project \citep[COUP;][]{2005ApJS..160..319G}, and the XMM-Newton Extended Survey of Taurus \citep[XEST;][]{2007AandA...468..353G}, look at entire young clusters, leveraging the ability to study many YSOs at once rather than focusing on a single source. Because of the multiple observations of XZ Tau AB over time discussed in Section \ref{sec:stable}, we can compare the ensemble of states of XZ Tau AB over time against other studies which have taken snapshots of an entire cluster at a given point in time.

To test this, we compared the temperatures and emission measures of our models of XZ Tau AB and those from previous work as a function of X-ray luminosity against the same characteristics for a subset of the data from COUP with two-temperature models that produce good fits (i.e.\ not meeting the ``marginal'' or ``poor'' flags in their paper), following the comparison of \citet{2005ApJS..160..401P} with the initial COUP dataset.

\begin{figure*}
\plottwo{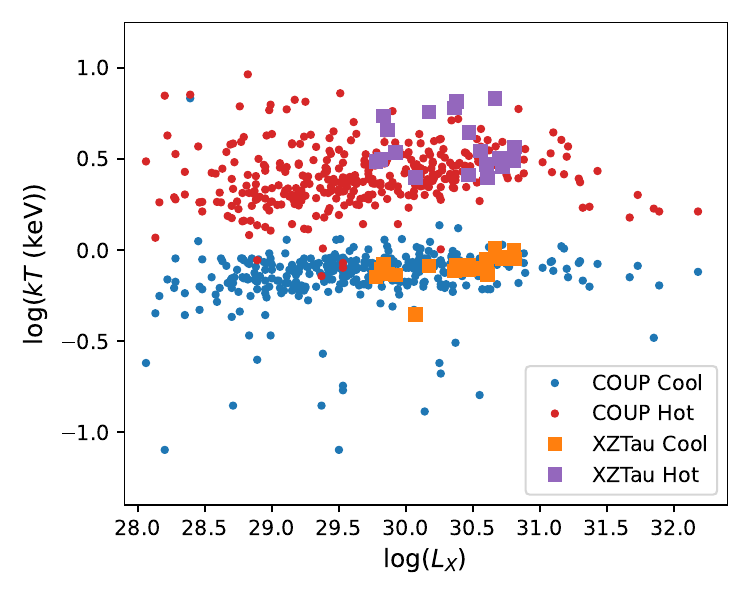}{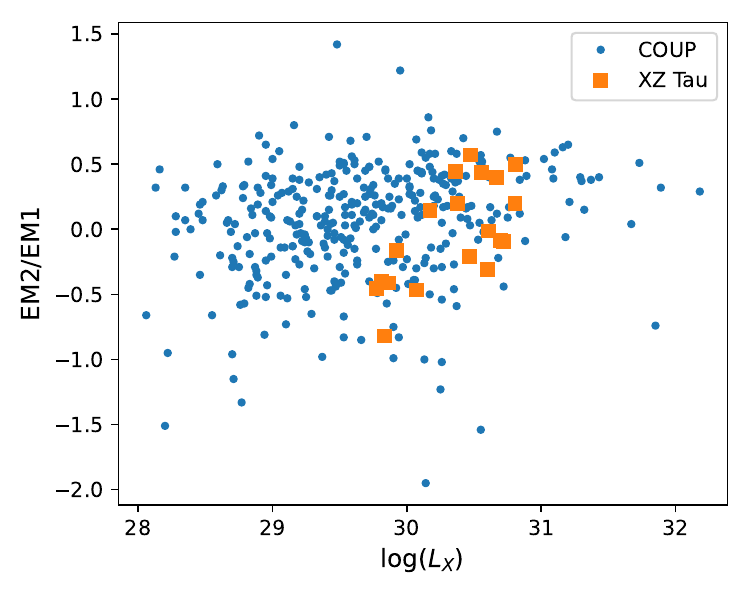}
\caption{Comparison of models of various observations of XZ Tau AB over time to model fits to the Chandra Orion Ultradeep Project (COUP). \textit{Left}: As with COUP model fits, XZ Tau AB over time exhibits a steady cool component and variable hot component that increase in brightness with increasing absorption-corrected X-ray luminosity. \textit{Right}: As with COUP model fits, the ratio of emission measures of warm and cool components increases (i.e.\ the warm component increases in emission) as a function of absorption-corrected X-ray luminosity.}
\label{fig:COUP_vs_XZTau}
\end{figure*}

As seen in Figure \ref{fig:COUP_vs_XZTau}, the XZ Tau AB data traces across the scatter in the COUP dataset. While the cool temperature component does stay steady as a function of unabsorbed X-ray luminosity, the hot component increases in temperature as a function of luminosity for both the COUP dataset and XZ Tau AB, an expected result of flare activity. XZ Tau AB over time similarly follows the trend across the COUP snapshot of increasing ratio of hot emission measure to cool as a function of X-ray luminosity (albeit with a fair amount of scatter). This suggests to us the possibility that much of the observed variation in the X-ray characteristics of YSOs is due primarily to the chance timing of the observation, rather than to fundamental differences in the stars themselves.

\section{Summary}
\label{sec:conclusion}

In this paper, we present CCD-resolution and gratings spectra of the pre-main sequence binary system XZ Tau AB from both \textit{Chandra} and \textit{XMM-Newton}. We find the following.

\begin{itemize}
    \item The \textit{Chandra} and \textit{XMM-Newton} data are both consistent with a two-temperature plasma-emission model.
    \item The cool component of the two-temperature model remains stable across both the \textit{Chandra} and \textit{XMM-Newton} observations, and is consistent with previous observations of the system with both observatories spanning more than a decade. This suggests a rotationally-induced magnetic field corona that remains stable over this timeframe, with recurring aperiodic flare activity superposed on top of this stable feature.
    \item The scatter in model fits of XZ Tau AB over time is comparable to the scatter in two-component model fits of the sources in COUP.
    \item Time-resolved spectra of two flares observed with \textit{XMM-Newton} show that the primary change during the flare is an increase in the volume emission measure of the ``hot'' component. Detailed analysis of the decay of one flare indicated a plasma loop length of $\sim 10 R_{\odot}$.
    \item XZ Tau AB is likely undergoing a long-term outburst typical of EXor objects during our X-ray observations based on the optical light curves. However, the change in X-ray brightness we see during several observations is more consistent with short-duration coronal flares than the long-term X-ray brightness increase seen in other EXOr objects \citep[e.g.\ V1647 Ori;][]{2012ApJ...754...32H}.
\end{itemize}

\begin{acknowledgments}
We thank the anonymous referee for insightful comments that were invaluable to the science presented in this work. S.M.S., S.J.W., and D.A.P. acknowledge support from the National Aeronautics and Space Administration through Chandra Award Number SAO GO8-19013X issued by the Chandra X-ray Observatory Center, which is operated by the Smithsonian Astrophysical Observatory for and on behalf of the National Aeronautics Space Administration under contract NAS8-03060. HMG acknowledges support from NASA grant 80NSSC21K0849. PCS acknowledges support from DLR 50OR2102.

This research has made use of data obtained from the Chandra Data Archive and the Chandra Source Catalog, and software provided by the Chandra X-ray Center (CXC) in the application packages CIAO and Sherpa.

Based on observations obtained with XMM-Newton, an ESA science mission with instruments and contributions directly funded by ESA Member States and NASA.

We acknowledge with thanks the variable star observations from the AAVSO International Database contributed by observers worldwide and used in this research.
\end{acknowledgments}


\vspace{5mm}
\facilities{CXO, XMM-Newton, AAVSO, ASAS-SN}


\software{AstroPy \citep{2013A&A...558A..33A,2018AJ....156..123A}, NumPy \citep{van2011numpy,harris2020array}, SciPy \citep{jones_scipy_2001,2020SciPy-NMeth}, Matplotlib \citep{Hunter:2007}, pandas \citep{mckinney,jeff_reback_2020_3644238}, CIAO \citep{2006SPIE.6270E..1VF}, Sherpa \citep{2001SPIE.4477...76F}}



\appendix

\section{Fitting New \textit{Chandra} and \textit{XMM-Newton} Data with Variable Abundances}

While we adopt the case that the Ne and Fe abundances are not significantly varying over the timeframe of our observations, we present here our best-fit models for these data, with Ne and Fe allowed to vary, for comparison.

\movetabledown=6.5cm
\begin{rotatetable*}
\begin{deluxetable*}{lccccccccccc}
\tablecaption{Summary of parameters for best fit models for X-ray spectra of XZ Tau AB with freely-varying Ne and Fe-like abundances \label{table:fitpars}}
\tablehead{
\colhead{} & \colhead{} & \colhead{$N_{H}$} & \multicolumn{4}{c}{Cool Component} & \multicolumn{2}{c}{Hot Component} & \multicolumn{2}{c}{Flux ($10^{-13}\, \mathrm{erg s^{-1}\,cm^{-2}}$)}  & \colhead{Log(Lum.)}\\
% (erg $\mathrm{s}^{-1}$)} 
\colhead{obsid} & \colhead{rstat\tablenotemark{a}} & \colhead{($10^{22} \mathrm{cm}^{-2}$)} & \colhead{kT (keV)} & \colhead{EM $10^{52} \mathrm{cm}^{-3}$} & \colhead{Ne ($\mathrm{Ne_{\odot}}$)} & \colhead{Fe ($\mathrm{Fe_{\odot}}$)} & \colhead{kT (keV)} & \colhead{EM ($10^{52} \mathrm{cm}^{-3}$)} & \colhead{Absorbed} & \colhead{Unabsorbed} & \colhead{log (erg $\mathrm{s}^{-1}$)}}             
\startdata
All \textit{Chandra} & $ 0.568 $ & $ 0.11                 $ & $ 0.83_{-0.6 }^{+0.1 } $ & $  1.0_{- 0.4}^{+  0.6} $ & $ 7.6_{-4.2}^{+ 4.8} $ & $ 0.9 _{-0.3 }^{+0.4 } $ & $ 2.2 _{-0.2 }^{+0.2 } $ & $  3.0 _{-0.4}^{+1.3} $ & $  2.5  $ & $  3.1  $ & 29.9 \\
201610   & $ 0.378 $ & $ 0.11                 $ & $ 0.29_{-0.3 }^{+1   } $ & $ 15  _{-14  }^{+649  } $ & $ 0.6_{-0.6}^{+31.6} $ & $ 0.9 _{-0.9 }^{+7.8 } $ & $ 3.2 _{-1.1 }^{+nan } $ & $  4   _{-3  }^{+2  } $ & $  7.4 $ & $ 11.6  $ & 30.5 \\
219510   & $ 0.131 $ & $ 0.11                 $ & $ 0.73_{-0.7 }^{+0.5 } $ & $  8  _{- 7  }^{+ 22  } $ & $ 2.6_{-2.6}^{+ 8.0} $ & $ 0.2 _{-0.2 }^{+0.9 } $ & $ 2.55_{nan  }^{+nan } $ & $  6   _{-5  }^{+7  } $ & $  5.7 $ & $  7.5  $ & 30.3 \\
Faint \textit{Chandra}\tablenotemark{b} & $ 0.769 $ & $ 0.11                 $ & $ 0.39_{-0.2 }^{+0.8 } $ & $  0.5_{- 0.3}^{+  0.6} $ & $ 7.4_{-3.8}^{+ 7.6} $ & $ 1.0 _{-0.4 }^{+0.7 } $ & $ 1.8 _{-0.2 }^{+0.2 } $ & $  2.6 _{-0.6}^{+0.6} $ & $  1.9 $ & $  2.4  $ & 29.8 \\
Faint \textit{XMM-Newton}\tablenotemark{c} & $ 0.915 $ & $ 0.11_{-0.01}^{+0.01} $ & $ 0.75_{-0.02}^{+0.02} $ & $  4.7_{- 0.6}^{+  0.6} $ & $ 1.4_{-0.2}^{+ 0.2} $ & $ 0.11_{-0.01}^{+0.02} $ & $ 3.2 _{-0.4 }^{+0.7 } $ & $  1.9 _{-0.3}^{+0.3} $ & $  2.1 $ & $  2.7  $ & 29.9 \\
0865040201      & $ 0.61  $ & $ 0.10_{-0.02}^{+0.02} $ & $ 0.74_{-0.05}^{+0.04} $ & $  5.1_{- 1.0}^{+  1.2} $ & $ 1.3_{-0.3}^{+ 0.3} $ & $ 0.10_{-0.02}^{+0.03} $ & $ 5.0 _{-1.2 }^{+3.8 } $ & $  2.1 _{-0.5}^{+0.4} $ & $  2.5  $ & $  3.1  $ & 29.9 \\
0865040301      & $ 0.573 $ & $ 0.08_{-0.01}^{+0.01} $ & $ 0.80_{-0.05}^{+0.05} $ & $  6.1_{- 1.2}^{+  1.1} $ & $ 1.6_{-0.3}^{+ 0.4} $ & $ 0.09_{-0.02}^{+0.02} $ & $ 5.5 _{-2.6 }^{+nan } $ & $  1.0 _{-0.4}^{+0.6} $ & $  2.3  $ & $  2.9  $ & 29.9 \\
0865040401      & $ 0.716 $ & $ 0.10_{-0.01}^{+0.01} $ & $ 0.85_{-0.02}^{+0.03} $ & $ 10.3_{- 1.3}^{+  1.6} $ & $ 1.8_{-0.3}^{+ 0.3} $ & $ 0.26_{-0.03}^{+0.04} $ & $ 2.44_{-0.09}^{+0.08} $ & $ 30.4 _{-0.8}^{+0.9} $ & $ 16.3 $ & $ 19.6  $ & 30.7 \\
0865040601      & $ 0.684 $ & $ 0.08_{-0.01}^{+0.01} $ & $ 0.78_{-0.04}^{+0.03} $ & $  6.0_{- 0.9}^{+  1.2} $ & $ 2.6_{-0.4}^{+ 0.4} $ & $ 0.26_{-0.04}^{+0.04} $ & $ 3.5 _{-0.1 }^{+0.1 } $ & $ 35.8 _{-0.8}^{+0.8} $ & $ 19.8 $ & $ 22.5  $ & 30.8 \\
0865040701      & $ 0.688 $ & $ 0.10_{-0.02}^{+0.02} $ & $ 0.79_{-0.05}^{+0.05} $ & $  4.2_{- 1.3}^{+  1.5} $ & $ 1.1_{-0.4}^{+ 0.5} $ & $ 0.13_{-0.03}^{+0.05} $ & $ 2.7 _{-0.5 }^{+1.1 } $ & $  2.3 _{-0.7}^{+0.6} $ & $  2.1 $ & $  2.7  $ & 29.8 \\
0865040501      & $ 0.619 $ & $ 0.15_{-0.02}^{+0.02} $ & $ 0.71_{-0.05}^{+0.04} $ & $  4.2_{- 1.1}^{+  1.1} $ & $ 1.3_{-0.3}^{+ 0.4} $ & $ 0.12_{-0.02}^{+0.04} $ & $ 2.9 _{-0.5 }^{+0.9 } $ & $  1.8 _{-0.4}^{+0.4} $ & $  1.7 $ & $  2.5  $ & 29.8 \\
\enddata
\tablenotetext{a}{Reduced $\chi^{2}$ using Gehrels weighting.}
\tablenotetext{b}{Joint fit of obsids 20160, 21946, 21947, 21948, 21950, 21952, 21953, 21954, and 21965.}
\tablenotetext{c}{Joint fit of obsids 0865040201, 0865040301, 0865040501, and 0865040701.}
\end{deluxetable*}
\end{rotatetable*}

\section{Summary of Parameters from Previous Observations}

We compiled fit parameters for previous observations of XZ Tau from \citet{2007AandA...468..353G,2007A&A...468..485F,2006A&A...453..241G,2020ApJ...888...15S}, in addition to our analysis here. While the methodologies between each paper are different enough to not be directly comparable on a numerical basis (e.g.\ different treatments of metallicity), they provide the backbone for the comparison presented in Figure \ref{fig:previous_work_vs_time}.

\begin{deluxetable*}{lccccccccccc}
\tablecaption{Comparison of Present Work to Previous Observations \label{tab:models_fluxes_previous}}
\tabletypesize{\footnotesize}
%\begin{deluxetable}{lccccccccccc}
\tablehead{ & \colhead{Abundance} & & & \colhead{Fit Bounds$^{b}$} & \colhead{$N_{H}\,^{c}$} & \multicolumn{2}{c}{kT (keV)} & \multicolumn{2}{c}{EM ($10^{52}\, \mathrm{{cm}^{-3}}$)} & \multicolumn{2}{c}{Flux ($10^{-13} \mathrm{erg\,cm^{-2}\,s^{-1}}$)} \\
\colhead{Source} & \colhead{Type} & \colhead{Uncertainty$^{a}$} & \colhead{Year} & \colhead{(keV)} & \colhead{($10^{22} \mathrm{cm^{-2}}$)} & \colhead{Cool} & \colhead{Hot} & \colhead{Cool} & \colhead{Hot} & \colhead{Absorbed} & \colhead{Unabsorbed}}
\startdata
XEST         & (1)    & 0.68 & 2000 & 0.5 - 7.3 & $0.24  _{-0.03 }^{+0.03 }$ & $0.7497              $ & $2.278                 $  & $ 4                        $ & $ 3.76                      $ & $2.078 $ & $3.38  $ \\
FranciosiniA & (1)    & 0.9  & 2000 & 0.3 - 7.3 & $0.245 _{-0.044}^{+0.061}$ & $0.73_{-0.11}^{+0.08}$ & $3.44_{ -0.95}^{+ 2.31}$ & $ 4.5  _{- 0.9  }^{+ 1.2  }$ & $ 3.1   _{-0.7   }^{+0.9   }$ & $2.217 $ & $3.567 $ \\
FranciosiniB & (1)    & 0.9  & 2000 & 0.3 - 7.3 & $0.205 _{-0.039}^{+0.047}$ & $0.82_{-0.09}^{+0.15}$ & $5.72_{ -1.71}^{+ 4.49}$ & $ 4.9  _{- 1.2  }^{+ 1.6  }$ & $ 6.8   _{-0.9   }^{+1.2   }$ & $4.735 $ & $6.324 $ \\
FranciosiniC & (1)    & 0.9  & 2000 & 0.3 - 7.3 & $0.191 _{-0.033}^{+0.039}$ & $0.77_{-0.12}^{+0.13}$ & $6.03_{ -1.51}^{+ 2.06}$ & $ 4.5  _{- 1.4  }^{+ 1.4  }$ & $12.5   _{-0.9   }^{+1.6   }$ & $7.775 $ & $9.783 $ \\
FranciosiniD & (1)    & 0.9  & 2000 & 0.3 - 7.3 & $0.222 _{-0.021}^{+0.023}$ & $0.78_{-0.09}^{+0.13}$ & $4.42_{ -0.45}^{+ 0.64}$ & $ 4.9  _{- 1.2  }^{+ 1.2  }$ & $18.3   _{-1.2   }^{+1.2   }$ & $9.662 $ & $12.63$ \\
FranciosiniE & (1)    & 0.9  & 2000 & 0.3 - 7.3 & $0.256 _{-0.026}^{+0.029}$ & $0.79_{-0.09}^{+0.09}$ & $3.5 _{ -0.42}^{+ 0.52}$ & $ 8.2  _{- 1.9  }^{+ 1.9  }$ & $22.3   _{-1.6   }^{+2.1   }$ & $10.82$ & $15.29$ \\
Giardino201  & (2)    & 0.68 & 2004 & 0.3 - 7.5 & $0.29  _{-0.06 }^{+0.06 }$ & $0.63_{-0.06}^{+0.06}$ & \nodata                  & $30.2  _{-13.1  }^{+13.1  }$ & \nodata                       & $0.7543$ & $6.388 $ \\
Giardino301  & (2)    & 0.68 & 2004 & 0.3 - 7.3 & $0.15  _{-0.03 }^{+0.03 }$ & $0.78_{-0.04}^{+0.04}$ & \nodata                  & $14.6  _{- 3.2  }^{+ 3.2  }$ & \nodata                       & $2.215 $ & $3.465 $ \\
Giardino401  & (2)    & 0.68 & 2004 & 0.3 - 7.3 & $0.08  _{-0.02 }^{+0.02 }$ & $0.77_{-0.04}^{+0.04}$ & \nodata                  & $12.5  _{- 2.6  }^{+ 2.6  }$ & \nodata                       & $2.294 $ & $2.952 $ \\
Giardino501  & (2)    & 0.68 & 2004 & 0.3 - 7.3 & $0.11  _{-0.04 }^{+0.04 }$ & $0.72_{-0.05}^{+0.05}$ & \nodata                  & $14.7  _{- 4.1  }^{+ 4.1  }$ & \nodata                       & $2.362 $ & $3.364 $ \\
Giardino601  & (2)    & 0.68 & 2004 & 0.3 - 7.3 & $0.07  _{-0.02 }^{+0.02 }$ & $0.78_{-0.04}^{+0.04}$ & \nodata                  & $11.9  _{- 5.2  }^{+ 0.4  }$ & \nodata                       & $2.265 $ & $2.824 $ \\
Giardino901  & (2)    & 0.68 & 2004 & 0.3 - 7.3 & $0.14  _{-0.03 }^{+0.03 }$ & $0.76_{-0.05}^{+0.05}$ & \nodata                  & $16.8  _{- 4.4  }^{+ 4.4  }$ & \nodata                       & $2.572 $ & $3.944 $ \\
Giardino1001 & (2)    & 0.68 & 2004 & 0.3 - 7.3 & $0.09  _{-0.02 }^{+0.02 }$ & $1.03_{-0.07}^{+0.07}$ & \nodata                  & $14.5  _{- 2.6  }^{+ 2.6  }$ & \nodata                       & $2.948 $ & $3.723 $ \\
Giardino1101 & (2)    & 0.68 & 2004 & 0.3 - 7.3 & $0.14  _{-0.02 }^{+0.02 }$ & $1.00_{-0.04}^{+0.04}$ & \nodata                  & $23.7  _{- 2.5  }^{+ 2.5  }$ & \nodata                       & $4.240 $ & $6.046 $ \\
Giardino1201 & (2)    & 0.68 & 2004 & 0.3 - 7.3 & $0.13  _{-0.02 }^{+0.02 }$ & $0.80_{-0.04}^{+0.04}$ & \nodata                  & $18    _{- 3.3  }^{+ 3.3  }$ & \nodata                       & $2.931 $ & $4.315 $ \\
Skinner      & (3)    & 0.68 & 2017 & 0.3 - 8   & $0.2                     $ & $0.26_{-0.03}^{+0.04}$ & $1.28_{ -0.04}^{+ 0.04}$ & $13.37 _{- 3.049}^{+ 3.752}$ & $13.84  _{-0.9381}^{+0.9381}$ & $3.446 $ & $5.895 $ \\
ChandraFull  & (4) & 0.68 & 2018 & 0.5 - 8   & $0.113                   $ & $0.49_{-0.05}^{+0.10}$ & $2.3 _{ -0.3 }^{+ 0.4 }$ & $ 4.6  _{- 0.6  }^{+ 0.7  }$ & $ 3.6   _{-0.7   }^{+0.5   }$ & $2.837 $ & $3.862 $ \\
FaintXMM     & (5)   & 0.68 & 2020 & 0.3 - 9   & $0.113 _{-0.007}^{+0.007}$ & $0.75_{-0.02}^{+0.02}$ & $3.2 _{ -0.3 }^{+ 0.5 }$ & $ 4.3  _{- 0.2  }^{+ 0.2  }$ & $ 1.7   _{-0.2   }^{+0.2   }$ & $2.078 $ & $2.739 $ \\
0865040201   & (4) & 0.68 & 2020 & 0.3 - 9   & $0.09  _{-0.01 }^{+0.01 }$ & $0.72_{-0.04}^{+0.04}$ & $4.6 _{ -0.9 }^{+ 1.6 }$ & $ 4.3  _{- 0.3  }^{+ 0.3  }$ & $ 2.0   _{-0.3   }^{+0.3   }$ & $2.477 $ & $3.06  $ \\
0865040301   & (4) & 0.68 & 2020 & 0.3 - 9   & $0.077 _{-0.009}^{+0.009}$ & $0.84_{-0.03}^{+0.03}$ & $5   _{ -2   }^{+14   }$ & $ 5.4  _{- 0.4  }^{+ 0.3  }$ & $ 0.9   _{-0.2   }^{+0.3   }$ & $2.326 $ & $2.831 $ \\
0865040401   & (4) & 0.68 & 2020 & 0.3 - 9   & $0.127 _{-0.005}^{+0.005}$ & $0.87_{-0.02}^{+0.02}$ & $2.8 _{ -0.1 }^{+ 0.2 }$ & $21    _{- 1    }^{+ 1    }$ & $23     _{-1     }^{+1     }$ & $16.31 $ & $20.84 $ \\
0865040401A  & (4) & 0.68 & 2020 & 0.3 - 9   & $0.127                   $ & $0.95_{-0.05}^{+0.06}$ & $3.0 _{ -0.2 }^{+ 0.3 }$ & $21    _{- 3    }^{+ 3    }$ & $39     _{-3     }^{+3     }$ & $23.98 $ & $28.71 $ \\
0865040401B  & (4) & 0.68 & 2020 & 0.3 - 9   & $0.127                   $ & $0.89_{-0.04}^{+0.04}$ & $2.9 _{ -0.3 }^{+ 0.4 }$ & $25    _{- 2    }^{+ 2    }$ & $23     _{-2     }^{+2     }$ & $17.92 $ & $22.08 $ \\
0865040401C  & (4) & 0.68 & 2020 & 0.3 - 9   & $0.127                   $ & $0.87_{-0.04}^{+0.04}$ & $3.1 _{ -0.6 }^{+ 1   }$ & $25    _{- 3    }^{+ 2    }$ & $13     _{-3     }^{+3     }$ & $13.92 $ & $17.49 $ \\
0865040401D  & (4) & 0.68 & 2020 & 0.3 - 9   & $0.127                   $ & $0.81_{-0.03}^{+0.04}$ & $2.6 _{ -0.4 }^{+ 0.5 }$ & $17    _{- 1    }^{+ 2    }$ & $11     _{-2     }^{+2     }$ & $9.909 $ & $12.54 $ \\
0865040601   & (4) & 0.68 & 2020 & 0.3 - 9   & $0.105 _{-0.005}^{+0.005}$ & $0.85_{-0.03}^{+0.03}$ & $3.8 _{ -0.2 }^{+ 0.2 }$ & $13.4  _{- 0.8  }^{+ 0.8  }$ & $31     _{-1     }^{+1     }$ & $19.76 $ & $23.27 $ \\
0865040601A  & (4) & 0.68 & 2020 & 0.3 - 9   & $0.105                   $ & $0.82_{-0.12}^{+0.10}$ & $9   _{ -3   }^{+ 9   }$ & $ 7    _{- 1    }^{+ 1    }$ & $11     _{-1     }^{+1     }$ & $9.291 $ & $10.73 $ \\
0865040601B  & (4) & 0.68 & 2020 & 0.3 - 9   & $0.105                   $ & $0.99_{-0.15}^{+0.14}$ & $7   _{ -2   }^{+ 6   }$ & $11    _{- 3    }^{+ 4    }$ & $25     _{-3     }^{+3     }$ & $18.65 $ & $21.23 $ \\
0865040601C  & (4) & 0.68 & 2020 & 0.3 - 9   & $0.105                   $ & $0.84_{-0.05}^{+0.05}$ & $3.8 _{ -0.2 }^{+ 0.2 }$ & $12    _{- 1    }^{+ 1    }$ & $42     _{-1     }^{+1     }$ & $25.03 $ & $29.13 $ \\
0865040601D  & (4) & 0.68 & 2020 & 0.3 - 9   & $0.105                   $ & $0.89_{-0.05}^{+0.05}$ & $3.0 _{ -0.4 }^{+ 0.6 }$ & $23    _{- 3    }^{+ 3    }$ & $22     _{-3     }^{+3     }$ & $17.32 $ & $21.27 $ \\
0865040601E  & (4) & 0.68 & 2020 & 0.3 - 9   & $0.105                   $ & $0.74_{-0.07}^{+0.06}$ & $2.4 _{ -0.4 }^{+ 0.5 }$ & $16    _{- 2    }^{+ 2    }$ & $20     _{-3     }^{+3     }$ & $13.07 $ & $16.38 $ \\
0865040701   & (4) & 0.68 & 2020 & 0.3 - 9   & $0.10  _{-0.01 }^{+0.02 }$ & $0.77_{-0.04}^{+0.04}$ & $2.9 _{ -0.5 }^{+ 0.8 }$ & $ 4.0  _{- 0.4  }^{+ 0.4  }$ & $ 2.0   _{-0.4   }^{+0.3   }$ & $2.146 $ & $2.752 $ \\
0865040501   & (4) & 0.68 & 2020 & 0.3 - 9   & $0.15  _{-0.01 }^{+0.02 }$ & $0.71_{-0.04}^{+0.04}$ & $2.9 _{ -0.5 }^{+ 0.7 }$ & $ 3.9  _{- 0.3  }^{+ 0.3  }$ & $ 1.6   _{-0.3   }^{+0.3   }$ & $1.724 $ & $2.497 $ \\
\enddata
%\end{deluxetable}
\tablecomments{(1) Fixed VAPEC model abundances described in \citet{2007AandA...468..353G}, based on \citet{2005ApJ...622..653T}, \citet{2004ApJ...609..925A}, \citet{2005ApJ...621.1009G}, and \citet{2005AandA...432..671S}. (2) Fixed metallicity of $Z=0.08Z_{\odot}$ based on average metallicity across fits to eleven individual observations. (3) Best fit metallicity of $Z=0.14_{-0.02}^{+0.03} Z_{\odot}$. (4) Fixed Ne and Fe values based on the best fit to a joint fit of the four \textit{XMM-Newton} observations not during flares. (5) Best-fit Ne and Fe values for this fit. \\
(a) Decimal uncertainties---i.e.~ 0.68 corresponds to 68\% uncertainties in the following numbers. \\
(b) Energy range considered when fitting---i.e. data outside these bounds are not considered in the fit. \\
(c) Values listed without uncertainties are fixed at the quoted value in the fitting.}
\end{deluxetable*}

\bibliography{references}{}

\begin{thebibliography}{}
\expandafter\ifx\csname natexlab\endcsname\relax\def\natexlab#1{#1}\fi
\providecommand{\url}[1]{\href{#1}{#1}}
\providecommand{\dodoi}[1]{doi:~\href{http://doi.org/#1}{\nolinkurl{#1}}}
\providecommand{\doeprint}[1]{\href{http://ascl.net/#1}{\nolinkurl{http://ascl.net/#1}}}
\providecommand{\doarXiv}[1]{\href{https://arxiv.org/abs/#1}{\nolinkurl{https://arxiv.org/abs/#1}}}

\bibitem[{{Arce} \& {Sargent}(2006)}]{2006ApJ...646.1070A}
{Arce}, H.~G., \& {Sargent}, A.~I. 2006, \apj, 646, 1070,
  \dodoi{10.1086/505104}

\bibitem[{{Argiroffi} {et~al.}(2004){Argiroffi}, {Drake}, {Maggio}, {Peres},
  {Sciortino}, \& {Harnden}}]{2004ApJ...609..925A}
{Argiroffi}, C., {Drake}, J.~J., {Maggio}, A., {et~al.} 2004, \apj, 609, 925,
  \dodoi{10.1086/420692}

\bibitem[{{Astropy Collaboration} {et~al.}(2013){Astropy Collaboration},
  {Robitaille}, {Tollerud}, {Greenfield}, {Droettboom}, {Bray}, {Aldcroft},
  {Davis}, {Ginsburg}, {Price-Whelan}, {Kerzendorf}, {Conley}, {Crighton},
  {Barbary}, {Muna}, {Ferguson}, {Grollier}, {Parikh}, {Nair}, {Unther},
  {Deil}, {Woillez}, {Conseil}, {Kramer}, {Turner}, {Singer}, {Fox}, {Weaver},
  {Zabalza}, {Edwards}, {Azalee Bostroem}, {Burke}, {Casey}, {Crawford},
  {Dencheva}, {Ely}, {Jenness}, {Labrie}, {Lim}, {Pierfederici}, {Pontzen},
  {Ptak}, {Refsdal}, {Servillat}, \& {Streicher}}]{2013A&A...558A..33A}
{Astropy Collaboration}, {Robitaille}, T.~P., {Tollerud}, E.~J., {et~al.} 2013,
  A\&A, 558, A33, \dodoi{10.1051/0004-6361/201322068}

\bibitem[{{Astropy Collaboration} {et~al.}(2018){Astropy Collaboration},
  {Price-Whelan}, {Sip{\H{o}}cz}, {G{\"u}nther}, {Lim}, {Crawford}, {Conseil},
  {Shupe}, {Craig}, {Dencheva}, {Ginsburg}, {VanderPlas}, {Bradley},
  {P{\'e}rez-Su{\'a}rez}, {de Val-Borro}, {Aldcroft}, {Cruz}, {Robitaille},
  {Tollerud}, {Ardelean}, {Babej}, {Bach}, {Bachetti}, {Bakanov}, {Bamford},
  {Barentsen}, {Barmby}, {Baumbach}, {Berry}, {Biscani}, {Boquien}, {Bostroem},
  {Bouma}, {Brammer}, {Bray}, {Breytenbach}, {Buddelmeijer}, {Burke},
  {Calderone}, {Cano Rodr{\'\i}guez}, {Cara}, {Cardoso}, {Cheedella}, {Copin},
  {Corrales}, {Crichton}, {D'Avella}, {Deil}, {Depagne}, {Dietrich}, {Donath},
  {Droettboom}, {Earl}, {Erben}, {Fabbro}, {Ferreira}, {Finethy}, {Fox},
  {Garrison}, {Gibbons}, {Goldstein}, {Gommers}, {Greco}, {Greenfield},
  {Groener}, {Grollier}, {Hagen}, {Hirst}, {Homeier}, {Horton}, {Hosseinzadeh},
  {Hu}, {Hunkeler}, {Ivezi{\'c}}, {Jain}, {Jenness}, {Kanarek}, {Kendrew},
  {Kern}, {Kerzendorf}, {Khvalko}, {King}, {Kirkby}, {Kulkarni}, {Kumar},
  {Lee}, {Lenz}, {Littlefair}, {Ma}, {Macleod}, {Mastropietro}, {McCully},
  {Montagnac}, {Morris}, {Mueller}, {Mumford}, {Muna}, {Murphy}, {Nelson},
  {Nguyen}, {Ninan}, {N{\"o}the}, {Ogaz}, {Oh}, {Parejko}, {Parley}, {Pascual},
  {Patil}, {Patil}, {Plunkett}, {Prochaska}, {Rastogi}, {Reddy Janga},
  {Sabater}, {Sakurikar}, {Seifert}, {Sherbert}, {Sherwood-Taylor}, {Shih},
  {Sick}, {Silbiger}, {Singanamalla}, {Singer}, {Sladen}, {Sooley},
  {Sornarajah}, {Streicher}, {Teuben}, {Thomas}, {Tremblay}, {Turner},
  {Terr{\'o}n}, {van Kerkwijk}, {de la Vega}, {Watkins}, {Weaver}, {Whitmore},
  {Woillez}, {Zabalza}, \& {Astropy Contributors}}]{2018AJ....156..123A}
{Astropy Collaboration}, {Price-Whelan}, A.~M., {Sip{\H{o}}cz}, B.~M., {et~al.}
  2018, \aj, 156, 123, \dodoi{10.3847/1538-3881/aabc4f}

\bibitem[{{Audard} {et~al.}(2014){Audard}, {{\'A}brah{\'a}m}, {Dunham},
  {Green}, {Grosso}, {Hamaguchi}, {Kastner}, {K{\'o}sp{\'a}l}, {Lodato},
  {Romanova}, {Skinner}, {Vorobyov}, \& {Zhu}}]{2014prpl.conf..387A}
{Audard}, M., {{\'A}brah{\'a}m}, P., {Dunham}, M.~M., {et~al.} 2014, in
  Protostars and Planets VI, ed. H.~{Beuther}, R.~S. {Klessen}, C.~P.
  {Dullemond}, \& T.~{Henning}, 387,
  \dodoi{10.2458/azu\_uapress\_9780816531240-ch017}

\bibitem[{{Bonito} {et~al.}(2007){Bonito}, {Orlando}, {Peres}, {Favata}, \&
  {Rosner}}]{2007A&A...462..645B}
{Bonito}, R., {Orlando}, S., {Peres}, G., {Favata}, F., \& {Rosner}, R. 2007,
  \aap, 462, 645, \dodoi{10.1051/0004-6361:20065236}

\bibitem[{{Brickhouse} {et~al.}(2010){Brickhouse}, {Cranmer}, {Dupree}, {Luna},
  \& {Wolk}}]{2010ApJ...710.1835B}
{Brickhouse}, N.~S., {Cranmer}, S.~R., {Dupree}, A.~K., {Luna}, G.~J.~M., \&
  {Wolk}, S. 2010, \apj, 710, 1835, \dodoi{10.1088/0004-637X/710/2/1835}

\bibitem[{{Canizares} {et~al.}(2005){Canizares}, {Davis}, {Dewey}, {Flanagan},
  {Galton}, {Huenemoerder}, {Ishibashi}, {Markert}, {Marshall}, {McGuirk},
  {Schattenburg}, {Schulz}, {Smith}, \& {Wise}}]{2005PASP..117.1144C}
{Canizares}, C.~R., {Davis}, J.~E., {Dewey}, D., {et~al.} 2005, \pasp, 117,
  1144, \dodoi{10.1086/432898}

\bibitem[{{Clarke} {et~al.}(1990){Clarke}, {Lin}, \&
  {Pringle}}]{1990MNRAS.242..439C}
{Clarke}, C.~J., {Lin}, D.~N.~C., \& {Pringle}, J.~E. 1990, \mnras, 242, 439,
  \dodoi{10.1093/mnras/242.3.439}

\bibitem[{{Cleeves} {et~al.}(2013){Cleeves}, {Adams}, \&
  {Bergin}}]{2013ApJ...772....5C}
{Cleeves}, L.~I., {Adams}, F.~C., \& {Bergin}, E.~A. 2013, \apj, 772, 5,
  \dodoi{10.1088/0004-637X/772/1/5}

\bibitem[{{Coffey} {et~al.}(2004){Coffey}, {Downes}, \&
  {Ray}}]{2004A&A...419..593C}
{Coffey}, D., {Downes}, T.~P., \& {Ray}, T.~P. 2004, \aap, 419, 593,
  \dodoi{10.1051/0004-6361:20034316}

\bibitem[{{Favata} {et~al.}(2005){Favata}, {Flaccomio}, {Reale}, {Micela},
  {Sciortino}, {Shang}, {Stassun}, \& {Feigelson}}]{2005ApJS..160..469F}
{Favata}, F., {Flaccomio}, E., {Reale}, F., {et~al.} 2005, \apjs, 160, 469,
  \dodoi{10.1086/432542}

\bibitem[{{Favata} {et~al.}(2002){Favata}, {Fridlund}, {Micela}, {Sciortino},
  \& {Kaas}}]{Favata_2002}
{Favata}, F., {Fridlund}, C.~V.~M., {Micela}, G., {Sciortino}, S., \& {Kaas},
  A.~A. 2002, \aap, 386, 204, \dodoi{10.1051/0004-6361:20011387}

\bibitem[{{Favata} {et~al.}(2003){Favata}, {Giardino}, {Micela}, {Sciortino},
  \& {Damiani}}]{2003A&A...403..187F}
{Favata}, F., {Giardino}, G., {Micela}, G., {Sciortino}, S., \& {Damiani}, F.
  2003, \aap, 403, 187, \dodoi{10.1051/0004-6361:20030305}

\bibitem[{{Foster} {et~al.}(2012){Foster}, {Ji}, {Smith}, \&
  {Brickhouse}}]{2012ApJ...756..128F}
{Foster}, A.~R., {Ji}, L., {Smith}, R.~K., \& {Brickhouse}, N.~S. 2012, \apj,
  756, 128, \dodoi{10.1088/0004-637X/756/2/128}

\bibitem[{{Franciosini} {et~al.}(2007){Franciosini}, {Pillitteri}, {Stelzer},
  {Micela}, {Briggs}, {Scelsi}, {Telleschi}, {Audard}, {Palla}, \&
  {G{\"u}del}}]{2007A&A...468..485F}
{Franciosini}, E., {Pillitteri}, I., {Stelzer}, B., {et~al.} 2007, \aap, 468,
  485, \dodoi{10.1051/0004-6361:20066536}

\bibitem[{{Freeman} {et~al.}(2001){Freeman}, {Doe}, \&
  {Siemiginowska}}]{2001SPIE.4477...76F}
{Freeman}, P., {Doe}, S., \& {Siemiginowska}, A. 2001, in Society of
  Photo-Optical Instrumentation Engineers (SPIE) Conference Series, Vol. 4477,
  Astronomical Data Analysis, ed. J.-L. {Starck} \& F.~D. {Murtagh}, 76--87,
  \dodoi{10.1117/12.447161}

\bibitem[{{Fruscione} {et~al.}(2006){Fruscione}, {McDowell}, {Allen},
  {Brickhouse}, {Burke}, {Davis}, {Durham}, {Elvis}, {Galle}, {Harris},
  {Huenemoerder}, {Houck}, {Ishibashi}, {Karovska}, {Nicastro}, {Noble},
  {Nowak}, {Primini}, {Siemiginowska}, {Smith}, \&
  {Wise}}]{2006SPIE.6270E..1VF}
{Fruscione}, A., {McDowell}, J.~C., {Allen}, G.~E., {et~al.} 2006, in Society
  of Photo-Optical Instrumentation Engineers (SPIE) Conference Series, Vol.
  6270, Society of Photo-Optical Instrumentation Engineers (SPIE) Conference
  Series, ed. D.~R. {Silva} \& R.~E. {Doxsey}, 62701V,
  \dodoi{10.1117/12.671760}

\bibitem[{{Garc{\'\i}a-Alvarez} {et~al.}(2005){Garc{\'\i}a-Alvarez}, {Drake},
  {Lin}, {Kashyap}, \& {Ball}}]{2005ApJ...621.1009G}
{Garc{\'\i}a-Alvarez}, D., {Drake}, J.~J., {Lin}, L., {Kashyap}, V.~L., \&
  {Ball}, B. 2005, \apj, 621, 1009, \dodoi{10.1086/427721}

\bibitem[{{Getman} \& {Feigelson}(2021)}]{2021ApJ...916...32G}
{Getman}, K.~V., \& {Feigelson}, E.~D. 2021, \apj, 916, 32,
  \dodoi{10.3847/1538-4357/ac00be}

\bibitem[{{Getman} {et~al.}(2021){Getman}, {Feigelson}, \&
  {Garmire}}]{2021ApJ...920..154G}
{Getman}, K.~V., {Feigelson}, E.~D., \& {Garmire}, G.~P. 2021, \apj, 920, 154,
  \dodoi{10.3847/1538-4357/ac1746}

\bibitem[{{Getman} {et~al.}(2005){Getman}, {Flaccomio}, {Broos}, {Grosso},
  {Tsujimoto}, {Townsley}, {Garmire}, {Kastner}, {Li}, {Harnden}, {Wolk},
  {Murray}, {Lada}, {Muench}, {McCaughrean}, {Meeus}, {Damiani}, {Micela},
  {Sciortino}, {Bally}, {Hillenbrand}, {Herbst}, {Preibisch}, \&
  {Feigelson}}]{2005ApJS..160..319G}
{Getman}, K.~V., {Flaccomio}, E., {Broos}, P.~S., {et~al.} 2005, \apjs, 160,
  319, \dodoi{10.1086/432092}

\bibitem[{{Giardino} {et~al.}(2006){Giardino}, {Favata}, {Silva}, {Micela},
  {Reale}, \& {Sciortino}}]{2006A&A...453..241G}
{Giardino}, G., {Favata}, F., {Silva}, B., {et~al.} 2006, \aap, 453, 241,
  \dodoi{10.1051/0004-6361:20053663}

\bibitem[{{Grosso} {et~al.}(2006){Grosso}, {Feigelson}, {Getman}, {Kastner},
  {Bally}, \& {McCaughrean}}]{2006A&A...448L..29G}
{Grosso}, N., {Feigelson}, E.~D., {Getman}, K.~V., {et~al.} 2006, \aap, 448,
  L29, \dodoi{10.1051/0004-6361:200600004}

\bibitem[{{G{\"u}del} {et~al.}(2008){G{\"u}del}, {Skinner}, {Audard}, {Briggs},
  \& {Cabrit}}]{2008A&A...478..797G}
{G{\"u}del}, M., {Skinner}, S.~L., {Audard}, M., {Briggs}, K.~R., \& {Cabrit},
  S. 2008, \aap, 478, 797, \dodoi{10.1051/0004-6361:20078141}

\bibitem[{{G{\"u}del} {et~al.}(2007){G{\"u}del}, {Briggs}, {Arzner}, {Audard},
  {Bouvier}, {Feigelson}, {Franciosini}, {Glauser}, {Grosso}, {Micela},
  {Monin}, {Montmerle}, {Padgett}, {Palla}, {Pillitteri}, {Rebull}, {Scelsi},
  {Silva}, {Skinner}, {Stelzer}, \& {Telleschi}}]{2007AandA...468..353G}
{G{\"u}del}, M., {Briggs}, K.~R., {Arzner}, K., {et~al.} 2007, A\&A, 468, 353,
  \dodoi{10.1051/0004-6361:20065724}

\bibitem[{{Haas} {et~al.}(1990){Haas}, {Leinert}, \&
  {Zinnecker}}]{1990A&A...230L...1H}
{Haas}, M., {Leinert}, C., \& {Zinnecker}, H. 1990, \aap, 230, L1

\bibitem[{{Hamaguchi} {et~al.}(2012){Hamaguchi}, {Grosso}, {Kastner},
  {Weintraub}, {Richmond}, {Petre}, {Teets}, \&
  {Principe}}]{2012ApJ...754...32H}
{Hamaguchi}, K., {Grosso}, N., {Kastner}, J.~H., {et~al.} 2012, \apj, 754, 32,
  \dodoi{10.1088/0004-637X/754/1/32}

\bibitem[{Harris {et~al.}(2020)Harris, Millman, van~der Walt, Gommers,
  Virtanen, Cournapeau, Wieser, Taylor, Berg, Smith, Kern, Picus, Hoyer, van
  Kerkwijk, Brett, Haldane, del R{'{\i}}o, Wiebe, Peterson,
  G{'{e}}rard-Marchant, Sheppard, Reddy, Weckesser, Abbasi, Gohlke, \&
  Oliphant}]{harris2020array}
Harris, C.~R., Millman, K.~J., van~der Walt, S.~J., {et~al.} 2020, Nature, 585,
  357, \dodoi{10.1038/s41586-020-2649-2}

\bibitem[{{Huenemoerder} {et~al.}(2007){Huenemoerder}, {Kastner}, {Testa},
  {Schulz}, \& {Weintraub}}]{2007ApJ...671..592H}
{Huenemoerder}, D.~P., {Kastner}, J.~H., {Testa}, P., {Schulz}, N.~S., \&
  {Weintraub}, D.~A. 2007, \apj, 671, 592, \dodoi{10.1086/522921}

\bibitem[{Hunter(2007)}]{Hunter:2007}
Hunter, J.~D. 2007, Computing In Science \& Engineering, 9, 90,
  \dodoi{10.1109/MCSE.2007.55}

\bibitem[{{Ichikawa} {et~al.}(2021){Ichikawa}, {Kido}, {Takaishi}, {Shimajiri},
  {Tsukamoto}, \& {Takakuwa}}]{2021ApJ...919...55I}
{Ichikawa}, T., {Kido}, M., {Takaishi}, D., {et~al.} 2021, \apj, 919, 55,
  \dodoi{10.3847/1538-4357/ac0dc3}

\bibitem[{{Jayasinghe} {et~al.}(2019){Jayasinghe}, {Stanek}, {Kochanek},
  {Shappee}, {Holoien}, {Thompson}, {Prieto}, {Dong}, {Pawlak}, {Pejcha},
  {Shields}, {Pojmanski}, {Otero}, {Britt}, \& {Will}}]{2019MNRAS.486.1907J}
{Jayasinghe}, T., {Stanek}, K.~Z., {Kochanek}, C.~S., {et~al.} 2019, \mnras,
  486, 1907, \dodoi{10.1093/mnras/stz844}

\bibitem[{Jones {et~al.}(2001)Jones, Oliphant, Peterson, \&
  Others}]{jones_scipy_2001}
Jones, E., Oliphant, T., Peterson, P., \& Others. 2001, {SciPy}: Open source
  scientific tools for Python.
\newblock \url{http://www.scipy.org/}

\bibitem[{{Kastner} {et~al.}(2002){Kastner}, {Huenemoerder}, {Schulz},
  {Canizares}, \& {Weintraub}}]{2002ApJ...567..434K}
{Kastner}, J.~H., {Huenemoerder}, D.~P., {Schulz}, N.~S., {Canizares}, C.~R.,
  \& {Weintraub}, D.~A. 2002, The Astrophysical Journal, 567, 434,
  \dodoi{10.1086/338419}

\bibitem[{{Kastner} {et~al.}(2004){Kastner}, {Richmond}, {Grosso}, {Weintraub},
  {Simon}, {Frank}, {Hamaguchi}, {Ozawa}, \& {Henden}}]{2004Natur.430..429K}
{Kastner}, J.~H., {Richmond}, M., {Grosso}, N., {et~al.} 2004, \nat, 430, 429,
  \dodoi{10.1038/nature02747}

\bibitem[{{Krist} {et~al.}(2008){Krist}, {Stapelfeldt}, {Hester}, {Healy},
  {Dwyer}, \& {Gardner}}]{2008AJ....136.1980K}
{Krist}, J.~E., {Stapelfeldt}, K.~R., {Hester}, J.~J., {et~al.} 2008, \aj, 136,
  1980, \dodoi{10.1088/0004-6256/136/5/1980}

\bibitem[{{Lorenzetti} {et~al.}(2009){Lorenzetti}, {Larionov}, {Giannini},
  {Arkharov}, {Antoniucci}, {Nisini}, \& {Di Paola}}]{2009ApJ...693.1056L}
{Lorenzetti}, D., {Larionov}, V.~M., {Giannini}, T., {et~al.} 2009, \apj, 693,
  1056, \dodoi{10.1088/0004-637X/693/2/1056}

\bibitem[{McKinney(2010)}]{mckinney}
McKinney, W. 2010, in Proceedings of the 9th Python in Science Conference, ed.
  S.~van~der Walt \& J.~Millman, 51 -- 56

\bibitem[{{Osorio} {et~al.}(2016){Osorio}, {Mac{\'\i}as}, {Anglada},
  {Carrasco-Gonz{\'a}lez}, {Galv{\'a}n-Madrid}, {Zapata}, {Calvet},
  {G{\'o}mez}, {Nagel}, {Rodr{\'\i}guez}, {Torrelles}, \&
  {Zhu}}]{2016ApJ...825L..10O}
{Osorio}, M., {Mac{\'\i}as}, E., {Anglada}, G., {et~al.} 2016, \apjl, 825, L10,
  \dodoi{10.3847/2041-8205/825/1/L10}

\bibitem[{{Owen} {et~al.}(2011){Owen}, {Ercolano}, \&
  {Clarke}}]{2011MNRAS.412...13O}
{Owen}, J.~E., {Ercolano}, B., \& {Clarke}, C.~J. 2011, \mnras, 412, 13,
  \dodoi{10.1111/j.1365-2966.2010.17818.x}

\bibitem[{{Pradhan} {et~al.}(2021){Pradhan}, {Huenemoerder}, {Ignace},
  {Pollock}, \& {Nichols}}]{2021ApJ...915..114P}
{Pradhan}, P., {Huenemoerder}, D.~P., {Ignace}, R., {Pollock}, A.~M.~T., \&
  {Nichols}, J.~S. 2021, \apj, 915, 114, \dodoi{10.3847/1538-4357/ac02c4}

\bibitem[{{Pravdo} {et~al.}(2001){Pravdo}, {Feigelson}, {Garmire}, {Maeda},
  {Tsuboi}, \& {Bally}}]{2001Natur.413..708P}
{Pravdo}, S.~H., {Feigelson}, E.~D., {Garmire}, G., {et~al.} 2001, \nat, 413,
  708, \dodoi{10.1038/35099508}

\bibitem[{{Preibisch} {et~al.}(2005){Preibisch}, {Kim}, {Favata}, {Feigelson},
  {Flaccomio}, {Getman}, {Micela}, {Sciortino}, {Stassun}, {Stelzer}, \&
  {Zinnecker}}]{2005ApJS..160..401P}
{Preibisch}, T., {Kim}, Y.-C., {Favata}, F., {et~al.} 2005, \apjs, 160, 401,
  \dodoi{10.1086/432891}

\bibitem[{{Reale} {et~al.}(2004){Reale}, {G{\"u}del}, {Peres}, \&
  {Audard}}]{2004A&A...416..733R}
{Reale}, F., {G{\"u}del}, M., {Peres}, G., \& {Audard}, M. 2004, \aap, 416,
  733, \dodoi{10.1051/0004-6361:20034027}

\bibitem[{Reback {et~al.}(2020)Reback, McKinney, jbrockmendel, den Bossche,
  Augspurger, Cloud, gfyoung, Sinhrks, Klein, Roeschke, Tratner, She, Hawkins,
  Ayd, Petersen, Schendel, Hayden, Garcia, MomIsBestFriend, Jancauskas,
  Battiston, Seabold, chris b1, h~vetinari, Hoyer, Overmeire, alimcmaster1,
  Mehyar, Dong, \& Whelan}]{jeff_reback_2020_3644238}
Reback, J., McKinney, W., jbrockmendel, {et~al.} 2020, pandas-dev/pandas:
  Pandas 1.0.1, v1.0.1,  Zenodo, \dodoi{10.5281/zenodo.3644238}

\bibitem[{{Scelsi} {et~al.}(2007){Scelsi}, {Maggio}, {Micela}, {Briggs}, \&
  {G{\"u}del}}]{2007A&A...473..589S}
{Scelsi}, L., {Maggio}, A., {Micela}, G., {Briggs}, K., \& {G{\"u}del}, M.
  2007, \aap, 473, 589, \dodoi{10.1051/0004-6361:20077792}

\bibitem[{{Scelsi} {et~al.}(2005){Scelsi}, {Maggio}, {Peres}, \&
  {Pallavicini}}]{2005AandA...432..671S}
{Scelsi}, L., {Maggio}, A., {Peres}, G., \& {Pallavicini}, R. 2005, \aap, 432,
  671, \dodoi{10.1051/0004-6361:20041739}

\bibitem[{{Schneider} {et~al.}(2011){Schneider}, {G{\"u}nther}, \&
  {Schmitt}}]{2011A&A...530A.123S}
{Schneider}, P.~C., {G{\"u}nther}, H.~M., \& {Schmitt}, J.~H.~M.~M. 2011, \aap,
  530, A123, \dodoi{10.1051/0004-6361/201016305}

\bibitem[{{Schneider} {et~al.}(2022){Schneider}, {G{\"u}nther}, \&
  {Ustamujic}}]{2022arXiv220706886S}
{Schneider}, P.~C., {G{\"u}nther}, H.~M., \& {Ustamujic}, S. 2022, arXiv
  e-prints, arXiv:2207.06886, \dodoi{10.48550/arXiv.2207.06886}

\bibitem[{{Schneider} \& {Schmitt}(2008)}]{2008A&A...488L..13S}
{Schneider}, P.~C., \& {Schmitt}, J.~H.~M.~M. 2008, \aap, 488, L13,
  \dodoi{10.1051/0004-6361:200810261}

\bibitem[{{Shappee} {et~al.}(2014){Shappee}, {Prieto}, {Grupe}, {Kochanek},
  {Stanek}, {De Rosa}, {Mathur}, {Zu}, {Peterson}, {Pogge}, {Komossa}, {Im},
  {Jencson}, {Holoien}, {Basu}, {Beacom}, {Szczygie{\l}}, {Brimacombe},
  {Adams}, {Campillay}, {Choi}, {Contreras}, {Dietrich}, {Dubberley},
  {Elphick}, {Foale}, {Giustini}, {Gonzalez}, {Hawkins}, {Howell}, {Hsiao},
  {Koss}, {Leighly}, {Morrell}, {Mudd}, {Mullins}, {Nugent}, {Parrent},
  {Phillips}, {Pojmanski}, {Rosing}, {Ross}, {Sand}, {Terndrup}, {Valenti},
  {Walker}, \& {Yoon}}]{2014ApJ...788...48S}
{Shappee}, B.~J., {Prieto}, J.~L., {Grupe}, D., {et~al.} 2014, \apj, 788, 48,
  \dodoi{10.1088/0004-637X/788/1/48}

\bibitem[{{Skinner} \& {G{\"u}del}(2013)}]{2013ApJ...765....3S}
{Skinner}, S.~L., \& {G{\"u}del}, M. 2013, \apj, 765, 3,
  \dodoi{10.1088/0004-637X/765/1/3}

\bibitem[{{Skinner} \& {G{\"u}del}(2020)}]{2020ApJ...888...15S}
---. 2020, \apj, 888, 15, \dodoi{10.3847/1538-4357/ab585c}

\bibitem[{{Stelzer} {et~al.}(2009){Stelzer}, {Hubrig}, {Orlando}, {Micela},
  {Mikul{\'a}{\v{s}}ek}, \& {Sch{\"o}ller}}]{2009A&A...499..529S}
{Stelzer}, B., {Hubrig}, S., {Orlando}, S., {et~al.} 2009, \aap, 499, 529,
  \dodoi{10.1051/0004-6361/200911750}

\bibitem[{{Telleschi} {et~al.}(2005){Telleschi}, {G{\"u}del}, {Briggs},
  {Audard}, {Ness}, \& {Skinner}}]{2005ApJ...622..653T}
{Telleschi}, A., {G{\"u}del}, M., {Briggs}, K., {et~al.} 2005, \apj, 622, 653,
  \dodoi{10.1086/428109}

\bibitem[{Van Der~Walt {et~al.}(2011)Van Der~Walt, Colbert, \&
  Varoquaux}]{van2011numpy}
Van Der~Walt, S., Colbert, S.~C., \& Varoquaux, G. 2011, Computing in Science
  \& Engineering, 13, 22

\bibitem[{Virtanen {et~al.}(2020)Virtanen, Gommers, Oliphant, Haberland, Reddy,
  Cournapeau, Burovski, Peterson, Weckesser, Bright, {van der Walt}, Brett,
  Wilson, Millman, Mayorov, Nelson, Jones, Kern, Larson, Carey, Polat, Feng,
  Moore, {VanderPlas}, Laxalde, Perktold, Cimrman, Henriksen, Quintero, Harris,
  Archibald, Ribeiro, Pedregosa, {van Mulbregt}, \& {SciPy 1.0
  Contributors}}]{2020SciPy-NMeth}
Virtanen, P., Gommers, R., Oliphant, T.~E., {et~al.} 2020, Nature Methods, 17,
  261, \dodoi{10.1038/s41592-019-0686-2}

\bibitem[{{Vorobyov}(2013)}]{2013A&A...552A.129V}
{Vorobyov}, E.~I. 2013, \aap, 552, A129, \dodoi{10.1051/0004-6361/201220601}

\bibitem[{{White} \& {Ghez}(2001)}]{2001ApJ...556..265W}
{White}, R.~J., \& {Ghez}, A.~M. 2001, \apj, 556, 265, \dodoi{10.1086/321542}

\bibitem[{{Zapata} {et~al.}(2015){Zapata}, {Galv{\'a}n-Madrid},
  {Carrasco-Gonz{\'a}lez}, {Curiel}, {Palau}, {Rodr{\'\i}guez}, {Kurtz},
  {Tafoya}, \& {Loinard}}]{2015ApJ...811L...4Z}
{Zapata}, L.~A., {Galv{\'a}n-Madrid}, R., {Carrasco-Gonz{\'a}lez}, C., {et~al.}
  2015, \apjl, 811, L4, \dodoi{10.1088/2041-8205/811/1/L4}

\end{thebibliography}
\bibliographystyle{aasjournal}



\end{document}